\documentclass[11pt,aps,prd,preprint,superscriptaddress]{revtex4-1}
\usepackage{amsmath}
\usepackage[dvips]{graphicx}
\usepackage[english]{babel}
\newcommand{\be}{\begin{equation}}
\newcommand{\ee}{\end{equation}}
\newcommand{\ben}{\begin{eqnarray}}
\newcommand{\een}{\end{eqnarray}}

\begin{document}
\title{A model of interacting holographic dark energy at the Ricci's scale}
\author{Iv\'{a}n Dur\'{a}n\footnote{E-mail: ivan.duran@uab.cat}}
\affiliation{Department of Physics, Autonomous University of
Barcelona, 08193 Bellaterra (Barcelona), Spain}
\author{Diego Pav\'{o}n\footnote{E-mail: diego.pavon@uab.es}}
\affiliation{Department of Physics, Autonomous University of
Barcelona, 08193 Bellaterra (Barcelona), Spain}
\begin{abstract}
We study a holographic cosmological model in which the infrared
cutoff is set by the Ricci's length and dark matter and dark
energy do not evolve on their own but interact non-gravitationally
with each other. This greatly alleviates the cosmic coincidence
problem because the ratio between both components does not vanish
at any time. We constrain the model with observational data from
supernovae, cosmic background radiation, baryon acoustic
oscillations, gas mass fraction in galaxy clusters, the history of
the Hubble function, and the growth function.
\end{abstract}

\maketitle

\section{Introduction}\label{introduction}
The efforts made to unveil the nature of dark energy (DE) -the
mysterious agent behind the present phase of cosmic accelerated
expansion- have been scarcely rewarded so far. We only know by
certain that DE is endowed with a hugely negative pressure (of the
order of its energy density) and we have grounds to suspect that
it is distributed rather evenly across space -see
\cite{recentreviews} to learn about the state of the art.

In view of its mysterious nature many authors have suggested that
dark energy should comply with the holographic principle which
asserts that the number of relevant degrees of freedom of a system
dominated by gravity must vary as the area of the surface bounding
the system \cite{gerard-leonard}. In addition, the energy density
of any given region should be bounded by that ascribed to a
Schwarzschild black hole that fills the same volume \cite{cohen}.
Mathematically this condition reads $\rho_{X}\leq M_{P}^{2}\,
L^{-2}$, where $\rho_{X}$ and $L$ stand for the DE density and the
size of the region (or infrared cutoff), respectively, and $M_{P}
= (8 \pi G)^{-1/2}$ is the reduced Planck mass. This expression is
most frequently written in its saturated form
%%%%%%%%%%%%%%%%%%%%%%%%%%%%%%%%%%%%%%%%%%%%%%%%%%%%%%%%%%%%%%%%%%%%%
\be \rho_{X}= \frac{3 M_{P}^{2}\,  c^{2}}{L^{2}}\, . \label{rhox}
\ee
%%%%%%%%%%%%%%%%%%%%%%%%%%%%%%%%%%%%%%%%%%%%%%%%%%%%%%%%%%%%%%%%%%%%%%
Here $c^{2}$ is a dimensionless parameter -very often, assumed
constant- that summarizes the uncertainties of the theory (such as
the number particle species and so on), and the factor $3$ was
introduced for mathematical convenience. An interesting feature of
holography lies in its close connection to the spacetime foam
\cite{jng}. For additional motivations of holographic dark energy
see section 3 of \cite{cqg-wd}. At any rate, it is sobering to
bear in  mind that the holographic proposal is just a reasonable
hypothesis (which we adopt in this paper) but not necessarily a
compelling one.

When dealing with holographic DE one must first specify the
infrared cutoff. In the lack of a clear guidance different
expressions have been adopted. The most relevant ones are the
Hubble radius, i.e., $L = H^{-1}$, see e.g. \cite{hubbleradius}-
and the Ricci's length, i.e., $L = (\dot{H} +2H^{2})^{-1/2}$ -see
e.g. \cite{gao,lxu,suwa,lepe}. The rationale behind the latter is
that it corresponds to the size of the maximal perturbation
leading to the formation of a black hole \cite{brustein}. The
radius of the future event horizon have been profusely used but it
suffers from a severe circularity problem.

The aim of this paper is to present a cosmological model of
holographic dark energy and constrain it with observational data.
The model takes the Ricci's length as infrared cutoff and assumes
that dark matter (DM) and dark energy do not evolve separately but
interact non-gravitationally with one another. The interaction,
albeit proposed just at a phenomenological level, is key to
alleviate the coincidence problem \cite{luca,srd} which cannot
find explanation in the $\Lambda$CDM model and afflicts so many
models of evolving DE. On the other hand, recently it has been
suggested that the dynamics of galaxy clusters can be better
explained if the said interaction is taken into account
\cite{abdalla-abramo}, and it has been invoked to explain the
nearly lack of correlation between the orientations of cluster
galaxy distributions with those of the underlying dark matter
distributions \cite{lee}. Further, the interaction -first proposed
to lower down the theoretical value of the cosmological constant
\cite{wetterich}- is not only necessary but inevitable
\cite{jerome}.

This paper is organized as follows. Section \ref{themodel}
introduces the model. Section \ref{discussion} discusses why
Ricci's holographic models seems to solve the coincidence problem
also in the absence of interaction. Section
\ref{statistical-analysis} presents the statistical analysis of
the model after constraining it with data from supernovae type Ia
(SN Ia), the shift of the first acoustic peak of the cosmic
background radiation (CMB-shift), baryon acoustic oscillations
(BAO), gas mass fraction in galaxy clusters, the history of the
Hubble parameter, $H(z)$, and the growth function. Lastly, section
\ref{conclusions} briefly delivers our main conclusions and offers
some final remarks. As usual, a zero subscript means the present
value of the corresponding quantity.
%%%%%%%%%%%%%%%%%%%%%%%%%%%%%%%%%%%%%%%%%%%%%%%%%%%%%%%%%%%%%%%%%%%%%%%%%%%%%%%%%%%%%%%%%%%%%%%
%%%%%%%%%%%%%%%%%%%%%%%%%%%%%%%%%%%%%%%%%%%%%%%%%%%%%%%%%%%%%%%%%%%%%%%%%%%%%%%%%%%%%%%%%%%%%%%
%%%%%%%%%%%%%%%%%%%%%%%%%%%%%%%%%%%%%%%%%%%%%%%%%%%%%%%%%%%%%%%%%%%%%%%%%%%%%%%%%%%%%%%%%%%%%%%
 \section{The holographic model}\label{themodel}
 This model assumes a spatially flat homogeneous and isotropic  universe dominated by
 DM and DE (subscripts $M$ and $X$, respectively), the latter obeying the holographic
 relationship (\ref{rhox}). In virtue of Friedmann equations,
%%%%%%%%%%%%%%%%%%%%%%%%%%%%%%%%%%%%%%%%%%%%%%%%%%%%%%%%%%%%%%%%%%%%%%%%%%%%%%%%%%%%
\begin{equation}\label{eq:friedmann1}
H^{2}  =  \frac{1}{3}M_{P}^{-2}(\rho_{M} \, + \, \rho_{X}) \, ,
\end{equation}
%%%%%%%%%%%%%%%%%%%%%%%%%%%%%%%%%%%%%%%%%%%%%%%%%%%%%%%%%%%%%%%%%%%%%%%%%%%%%%%%%%%%%%%
and
%%%%%%%%%%%%%%%%%%%%%%%%%%%%%%%%%%%%%%%%%%%%%%%%%%%%%%%%%%%%%%%%%%%%%%%%%%%%%%%%%%%%%%%
\begin{equation}\label{eq:friedmann2}
\dot{H}=  -\frac{1}{2}M_{P}^{-2}( \rho_{M} \, + \,
\rho_{X}+p_{X}) \, ,
\end{equation}
%%%%%%%%%%%%%%%%%%%%%%%%%%%%%%%%%%%%%%%%%%%%%%%%%%%%%%%%%%%%%%%%%%%%%%%%%%%%%%%%%%%%%%
 the fractional DE density, $\Omega_{X}=\frac{\rho_{X}}{3M_{P}^{2}H^{2}}$, can be
 expressed as
%%%%%%%%%%%%%%%%%%%%%%%%%%%%%%%%%%%%%%%%%%%%%%%%%%%%%%%%%%%%%%%
\begin{equation}\label{eq:Omega}
\Omega_{X}=\frac{c^{2}}{3c^{2}w+2} \, ,
\end{equation}
%%%%%%%%%%%%%%%%%%%%%%%%%%%%%%%%%%%%%%%%%%%%%%%%%%%%%%%%%%%%%%%%%%%%%%%
%%or equivalently,
%%%%%%%%%%%%%%%%%%%%%%%%%%%%%%%%%%%%%%%%%%%%%%%%%%%%%%%%%%%%%%%%%%%%%%%%%
%\begin{equation}\label{eq:w}
%w = \frac{1}{3}\left(\frac{1}{\Omega_{X}}-\frac{2}{c^{2}}\right),
%\end{equation}
%%%%%%%%%%%%%%%%%%%%%%%%%%%%%%%%%%%%%%%%%%%%%%%%%%%%%%%%%%%%%%%%%%%%%%%
where $w = p_{X}/\rho_{X}$ denotes the equation of state parameter
of dark energy which in general depends on time.

The deceleration parameter $q= - \ddot{a}/(a^{2} H^{2})$ takes the
simple expression,
%%%%%%%%%%%%%%%%%%%%%%%%%%%%%%%%%%%%%%%%%%%%%%%%%%%%%%%%%%%%%%%%%%%%%%%%%
\begin{equation}\label{eq:qDeOmega}
q = 1-\frac{\Omega_{X}}{c^{2}} \, .
\end{equation}
%%%%%%%%%%%%%%%%%%%%%%%%%%%%%%%%%%%%%%%%%%%%%%%%%%%%%%%%%%%%%%%%%%%%%%%%%%%%%%%%%%%
As we shall see, as a consequence of the evolution of
$\Omega_{X}$, it goes monotonously from positive values at early
times (in the matter dominated era) to negative values at later
times (in the dark energy dominated era).

The evolution of $\Omega_{X}$ follows from the conservation
equations of DE and DM. In the absence of  non-gravitational
interactions between them they evolve independently and obey
%%%%%%%%%%%%%%%%%%%%%%%%%%%%%%%%%%%%%%%%%%%%%%%%%%%%%%%%%%%%%%%%%%%%%%%%%%%%%%%%%%%%%%%
\begin{eqnarray}\label{eq:EvolNoIn}
\dot{\Omega}_{M}- \left(1 -\frac{2\Omega_{X}}{c^{2}}\right) (1-\Omega_{X})\, H&=&0 \, , \\
\label{eq:EvolNoIn2} \dot{\Omega}_{X}+ \left(1
-\frac{2\Omega_{X}}{c^{2}}\right)(1-\Omega_{X})\, H&=&0 \, .
\end{eqnarray}
%%%%%%%%%%%%%%%%%%%%%%%%%%%%%%%%%%%%%%%%%%%%%%%%%%%%%%%%%%%%%%%%%%%%%%%%%%%%%%%%%%%%%%

Bearing in mind that in our case  $\Omega_{M} \, + \, \Omega_{X} =
1$, we get the following expressions in terms of the redshift ($z
= a^{-1} -1$),
%%%%%%%%%%%%%%%%%%%%%%%%%%%%%%%%%%%%%%%%%%%%%%%%%%%%%%%%%%%%%%%%%%%%%%%%%%%%%%%%%%%%%%
\begin{equation}\label{eq:OmegaDez}
\Omega_{X}=\frac{2\Omega_{X0}-c^{2}+c^{2}(1-\Omega_{X0})(1+z)^{\frac{2}{c^{2}}-1}}
{2\Omega_{X0}-c^{2}+2(1-\Omega_{X0})(1+z)^{\frac{2}{c^{2}}-1}}\, .
\end{equation}
%%%%%%%%%%%%%%%%%%%%%%%%%%%%%%%%%%%%%%%%%%%%%%%%%%%%%%%%%%%%%%%%%%%%%%%%%%%%%%%%%%%%%%%%

A very useful quantity when considering the coincidence problem is
the ratio between the energy densities, $r = \rho_{M}/\rho_{X}$,
which for spatially flat universes reduces to
$\Omega_{M}/\Omega_{X}$. The coincidence problem gets alleviated
if for reasonable values of $\Omega_{X0}$ and $c^{2}$ we get
$r_{0} \sim {\cal O}(1)$. This is the case here since for
$\Omega_{X0}$ and $c^{2}$ of order unity, $r_{0}$ also results of
this same order. However, for late times (i.e, when $z \rightarrow
-1$) one has $r \rightarrow 0$. In this sense, the coincidence
problem is not properly solved. To obtain a non-vanishing ratio at
late times some interaction between DM and DE must be incorporated
in the picture \cite{prd-lad}.

If DM and DE interact non-gravitationally with each other the
evolution equations may be generalized as
%%%%%%%%%%%%%%%%%%%%%%%%%%%%%%%%%%%%%%%%%%%%%%%%%%%%%%%%%%%%%%%%%%%%%%%%%%%%%%%%%%%%%%%%%%
\begin{eqnarray}\label{eq:EvolIn}
\dot{\Omega}_{M} -  \left(1-\frac{2\Omega_{X}}{c^{2}}\right)(1-\Omega_{X})\, H&=&QH \, ,\\
\label{eq:EvolIn2} \dot{\Omega}_{X}+
\left(1-\frac{2\Omega_{X}}{c^{2}}\right)(1-\Omega_{X})\, H&=&-QH
\, ,
\end{eqnarray}
%%%%%%%%%%%%%%%%%%%%%%%%%%%%%%%%%%%%%%%%%%%%%%%%%%%%%%%%%%%%%%%%%%%%%%%%%%%%%%%%%%%%%%%%%%%
where the Hubble factor on the right hand sides has been
introduced to render the interaction term, $Q$, dimensionless.

Since the nature of  both dark components is largely unknown,
there is ample latitude in choosing $Q$. We shall specify it by
demanding that  $r$ evolves from an unstable fixed point in the
far past, $r_{\infty} \equiv r(z \rightarrow \infty)$, to a stable
fixed point at the far future, $r_{f} \equiv r(z \rightarrow -1)$
\cite{ladw}.
%%%%%%%%%%%%%%%%%%%%%%%%%%%%%%%%%%%%%%%%%%%%%%%%%%%%%%%%%%%%%%%%%%%%%%%%%%%%%%%%%%%%%%%%%%%
%%%%%%%%%%%%%%%%%%%%%%%%%%%%%%%%%%%%%%%%%%%%%%%%%%%%%%%%%%%%%%%%%%%%%%%%%%%%%%%%%%%%%%%%%%%

The pair of equations (\ref{eq:EvolIn}) and (\ref{eq:EvolIn2})
imply
%%%%%%%%%%%%%%%%%%%%%%%%%%%%%%%%%%%%%%%%%%%%%%%%%%%%%%%%%%%%%%%%%%%%%%%%%%%%%%%%%%%%%%%%%%%%%%%%%%%%%
\begin{equation}\label{eq:rDotChim}
\dot{r}= \left[r \left(1+r-\frac{2}{c^{2}}\right) +
Q(1+r)^{2}\right]H \, .
\end{equation}
%%%%%%%%%%%%%%%%%%%%%%%%%%%%%%%%%%%%%%%%%%%%%%%%%%%%%%%%%%%%%%%%%%%%%%%%%%%%%%%%%%%%%%%%%%%%%%%%%%%%%
\noindent Imposing that $r_{f}$ be a fixed point, i.e.,
$\dot{r}|_{r = r_{f}} = 0$ the interaction term $Q$ is simply a
constant given by
%%%%%%%%%%%%%%%%%%%%%%%%%%%%%%%%%%%%%%%%%%%%%%%%%%%%%%%%%%%%%%%%%%%%%%%%%%%%%%%%%%%%%%%%%%%%%%%%%%%%%
\begin{equation}\label{eq:QChim}
Q =
-\frac{r_{f}}{(1+r_{f})^{2}}\left(1+r_{f}-\frac{2}{c^{2}}\right)
\, .
\end{equation}
%%%%%%%%%%%%%%%%%%%%%%%%%%%%%%%%%%%%%%%%%%%%%%%%%%%%%%%%%%%%%%%%%%%%%%%%%%%%%%%%%%%%%%%%%%%%%%%%%%%%%
\noindent As we shall see later, $r_{f}$ and $c^{2}$ take values
such that $Q$ is positive-definite, which entails a transfer of
energy from dark energy to dark matter. Obviously if  $Q$ were
negative, the transfer of energy would go in the opposite
direction which would conflict with the second law of
thermodynamics \cite{db-grg} and the coincidence problem would
only worsen.

Rewriting Eq. (\ref{eq:rDotChim}) as
%%%%%%%%%%%%%%%%%%%%%%%%%%%%%%%%%%%%%%%%%%%%%%%%%%%%%%%%%%%%%%%%%%%%%%%%%%%%%%%%%%%%%%%%%%%%%%%%%%%%%
\begin{equation}\label{eq:rDotChim2}
\dot{r}= (Q+1)(r-r_{f})(r-r_{\infty})H
\end{equation}
%%%%%%%%%%%%%%%%%%%%%%%%%%%%%%%%%%%%%%%%%%%%%%%%%%%%%%%%%%%%%%%%%%%%%%%%%%%%%%%%%%%%%%%%%%%%%%%%%%%%%
\noindent and using the condition $\dot{r}=0$, the other fixed
point can be expressed in terms of the previous one, namely,
%%%%%%%%%%%%%%%%%%%%%%%%%%%%%%%%%%%%%%%%%%%%%%%%%%%%%%%%%%%%%%%%%%%%%%%%%%%%%%%%%%%%%%%%%%%%%%%%%%%%%
\begin{equation}
r_{\infty}= \frac{2-c^{2}(1+r_{f})}{2r_{f}+c^{2}(1+r_{f})}.
\label{eq:rfarpast}
\end{equation}
%%%%%%%%%%%%%%%%%%%%%%%%%%%%%%%%%%%%%%%%%%%%%%%%%%%%%%%%%%%%%%%%%%%%%%%%%%%%%%%%%%%%%%%%%%%%%%%%%%%%%

To study the stability of the  fixed points we first write $r{'}
\equiv dr/d\ln a = \dot{r}/H$ and calculate the derivative of
$r{'}$ with respect to $r$. In the case of the far future fixed
point we get
%%%%%%%%%%%%%%%%%%%%%%%%%%%%%%%%%%%%%%%%%%%%%%%%%%%%%%%%%%%%%%%%%%%%%%%%%%%%%%%%%%%%%%%%%%%%%%%%%%%%%
\begin{equation}\label{eq:fixedP1}
\frac{dr'}{dr}|_{r_{f}}= 1+\frac{2(r_{f}-1)}{c^{2}(r_{f}+1)} \, .
\end{equation}
%%%%%%%%%%%%%%%%%%%%%%%%%%%%%%%%%%%%%%%%%%%%%%%%%%%%%%%%%%%%%%%%%%%%%%%%%%%%%%%%%%%%%%%%%%%%%%%%%%%%%
\noindent Since $r_{f}$ must be lower than $r_{0} \simeq 0.45$,
from Eq. (\ref{eq:qDeOmega}) with  $c^{2} < \Omega_{X0} \simeq
0.75$ (otherwise $q_{0}$ would not be negative), one follows that
$\frac{dr'}{dr}|_{r_{f}} < 0 $, i.e., the fixed point $r_{f}$ is a
stable one. Similarly, we find that
%%%%%%%%%%%%%%%%%%%%%%%%%%%%%%%%%%%%%%%%%%%%%%%%%%%%%%%%%%%%%%%%%%%%%%%%%%%%%%%%%%%%%%%%%%%%%%%%%%%%%
\begin{equation}\label{eq:fixedP22}
\frac{dr'}{dr}|_{r_{\infty}}=-\frac{2+c^{2}}{c^{2}}+\frac{4}{c^{2}(1+r_{f})}
> 0 \, ,
\end{equation}
%%%%%%%%%%%%%%%%%%%%%%%%%%%%%%%%%%%%%%%%%%%%%%%%%%%%%%%%%%%%%%%%%%%%%%%%%%%%%%%%%%%%%%%%%%%%%%%%%%%%%
\noindent i.e., the fixed point at the far past is an unstable
one.

Equation (\ref{eq:rDotChim2}) can be integrated with the help of
(\ref{eq:QChim}). In terms of the redshift it yields,
%%%%%%%%%%%%%%%%%%%%%%%%%%%%%%%%%%%%%%%%%%%%%%%%%%%%%%%%%%%%%%%%%%%%%%%%%%%%%%%%%%%%%%%%%%%%%%%%%%%%
\begin{equation}\label{eq:rDezChim}
r = \frac{r_{f}(r_{0}-r_{\infty})-r_{\infty}(r_{0}-r_{f})(1+z)^{(Q
+1)(r_{\infty}-r_{f})}}{(r_{0}-r_{\infty})-(r_{0}-r_{f})(1+z)^{(Q+1)(r_{\infty}-r_{f})}}
\, .
\end{equation}
%%%%%%%%%%%%%%%%%%%%%%%%%%%%%%%%%%%%%%%%%%%%%%%%%%%%%%%%%%%%%%%%%%%%%%%%%%%%%%%%%%%%%%%%%%%%%%%%%%%%
%%%%%%%%%%%%%%%%%%%%%%%%%%%%%%%%%%%%%%%%%%%%%%%%%%%%%%%%%%%%%%%%%%%%%%%%%%%%%%%%%%%%%%%%%%%%%%%%%%%%
\begin{figure}[!hb]
  \begin{center}
    \begin{tabular}{c}
       \resizebox{100mm}{!}{\includegraphics{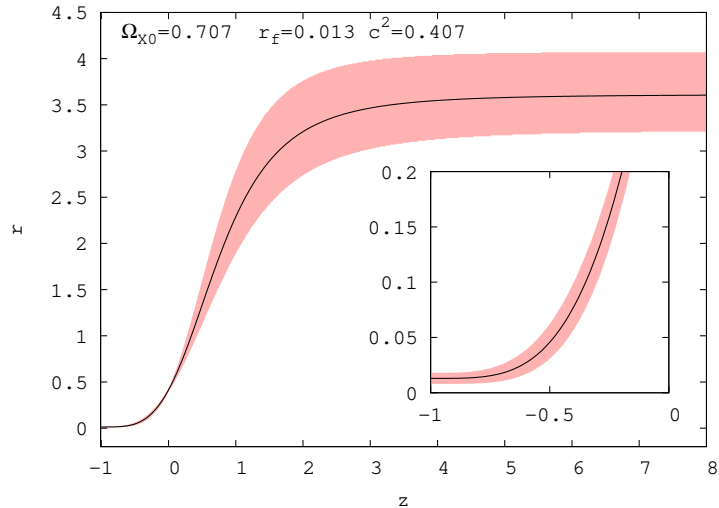}} \\
    \end{tabular}
    \caption{The ratio $r$ between the energy densities vs. redshift
    for the best fit model. As the inset shows $r_{f} \equiv r(z \rightarrow -1)$ does
    not vanish. In this, as well as in subsequent figures, the red swath indicates the
    region obtained by including the $1\sigma$ uncertainties of
    the constrained parameters used in the calculation.}
    \label{fig:r}
  \end{center}
\end{figure}

Inspection of (\ref{eq:rDotChim2}) readily shows that when $r$
lies between both fixed points one has $\dot{r} < 0$, i.e., the
ratio between the energy densities diminishes monotonously from
one fixed point to the other. This is depicted in Fig.
\ref{fig:r}. The said ratio smoothly decreases from high $z$
(i.e., from $r_{\infty}$ -the unstable fixed point is at $z
\rightarrow \infty)$) to asymptotically approach the fixed stable
point, $r_{f}$, at $z = -1$. Note that the latter needs not be
zero. In this regard the coincidence problem is much alleviated
because we are not living in any special era. However, the problem
is not solved in full since the model cannot predict that $r_{0}$
is of order unity. To the best of our knowledge, no model predicts
that, as well as no model predicts the present value of the
temperature of the cosmic background radiation, the Hubble
constant, or the age of the Universe. For the time being, we must
content ourselves by taking these values as input parameters
since, very possibly, we are to wait for a successful theory of
quantum gravity to compute them.

\begin{figure}[!htb]
  \begin{center}
    \begin{tabular}{cc}
       \resizebox{70mm}{!}{\includegraphics{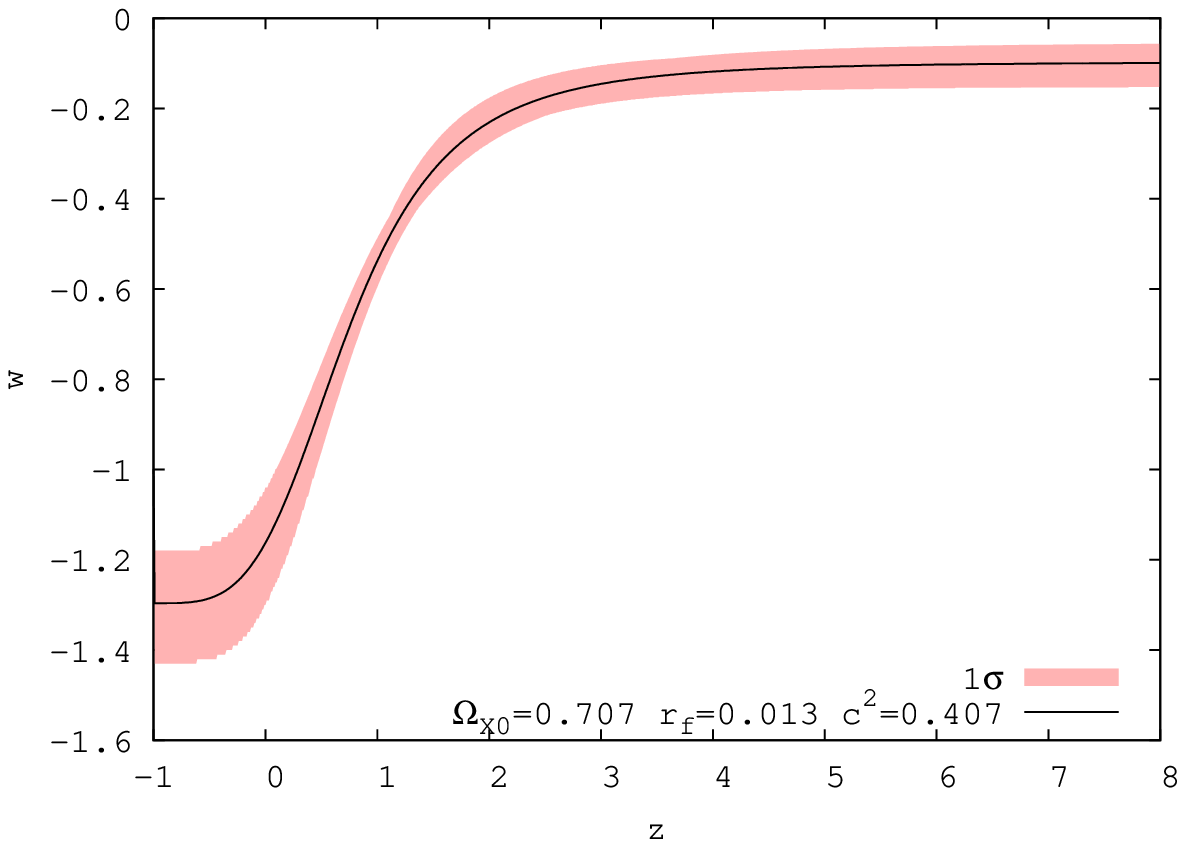}} &
       \resizebox{70mm}{!}{\includegraphics{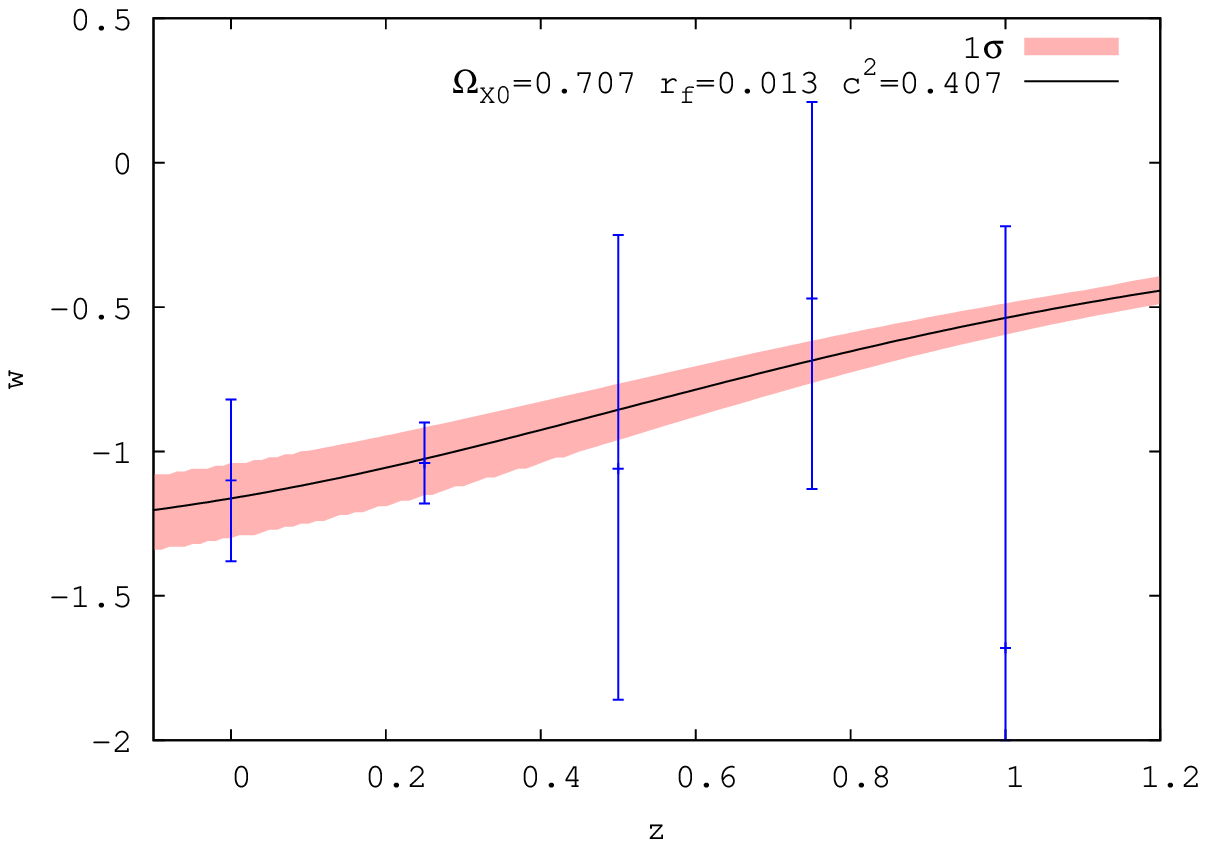}} \\
    \end{tabular}
    \caption{The equation of state parameter (as given
    by Eq. (\ref{eq:wDezChim})) vs. redshift up to $z = 8$ (left panel), and
    up to $ z = 1.2$ only (right panel) for the best fit holographic model. At
    high redshifts $w$ approaches the equation of state of non-relativistic matter
    and at low redshifts it does not depart significantly from $-1$. The observational
    data are taken from  \cite{serra-prd}; each error bar signifies a $2\sigma$ uncertainty.}
    \label{fig:w}
  \end{center}
\end{figure}

The expression for the fractional density of dark energy follows
from the relationship $r = (1-\Omega_{X})/\Omega_{X}$ and Eq.
(\ref{eq:rDezChim}),
%%%%%%%%%%%%%%%%%%%%%%%%%%%%%%%%%%%%%%%%%%%%%%%%%%%%%%%%%%%%%%%%%%%%%%%%%%%%%%%%%%%%%%%%%%%%%%%%%%%%%
\begin{equation}\label{eq:OmegaDezChim}
\Omega_{X}=\frac{(r_{0}-r_{\infty})-(r_{0}-r_{f})(1+z)^{(Q+1)(r_{\infty}-r_{f})}}
{(r_{f}+1)(r_{0}-r_{\infty})-(r_{\infty}+1)(r_{0}-r_{f})(1+z)^{(Q+1)(r_{\infty}-r_{f})}}\,
. \end{equation}
%%%%%%%%%%%%%%%%%%%%%%%%%%%%%%%%%%%%%%%%%%%%%%%%%%%%%%%%%%%%%%%%%%%%%%%%%%%%%%%%%%%%%%%%%%%%%%%%%%%%%
From the latter and (\ref{eq:Omega}) we obtain the equation of
state of dark energy in terms of the redshift,
%%%%%%%%%%%%%%%%%%%%%%%%%%%%%%%%%%%%%%%%%%%%%%%%%%%%%%%%%%%%%%%%%%%%%%%%%%%%%%%%%%%%%%%%%%%%%%%%%%%%%
\begin{equation}\label{eq:wDezChim}
w=\frac{1}{3}\left(1-\frac{2}{c^{2}}+r_{\infty}+\frac{(r_{0}-r_{\infty})(r_{\infty}-r_{f})}
{r_{\infty}-r_{0}+(r_{0}-r_{f})(1+z)^{(Q+1)(r_{\infty}-r_{f})}}\right)
\, .
\end{equation}
%%%%%%%%%%%%%%%%%%%%%%%%%%%%%%%%%%%%%%%%%%%%%%%%%%%%%%%%%%%%%%%%%%%%%%%%%%%%%%%%%%%%%%%%%%%%%%%%%%%%%
As shown in the left panel of Fig. \ref{fig:w} for the best fit
model, $w$ smoothly evolves from a negative value close to zero at
high redshifts to a value lower than $-1$ at the far future. The
right panel depicts its evolution near the present time ($z = 0$)
showing compatibility with recent observational data which suggest
that $w$ does not depart much from $-1$ at sufficiently low
redshifts (see \cite{serra-prd}).
%%%%%%%%%%%%%%%%%%%%%%%%%%%%%%%%%%%%%%%%%%%%%%%%%%%%%%%%%%%%%%%%%%%%%%%%%%%%%%%%%%%%%%%%%%%%%%%%%%%%%
%%%%%%%%%%%%%%%%%%%%%%%%%%%%%%%%%%%%%%%%%%%%%%%%%%%%%%%%%%%%%%%%%%%%%%%%%%%%%%%%%%%%%%%%%%%%%%%%%%%%%%

Integration of the second Friedmann's equation
(\ref{eq:friedmann2}) provides us with the evolution equation for
the Hubble factor which is key to perform the statistical analysis
of section \ref{statistical-analysis},
%%%%%%%%%%%%%%%%%%%%%%%%%%%%%%%%%%%%%%%%%%%%%%%%%%%%%%%%%%%%%%%%%%%%%%%%%%%%%%%%%%%%%%%%%%%%%%%%%%%%%%
\begin{equation}\label{eq:HDezChim}
H = H_{0} \left[\frac{A_{1}+2\left(A_{2}+
(r_{f}-r_{0})(1+z)^{-A_{3}}\right)}{A_{4}}
\right]^{1/2}(1+z)^{A_{5}} \, ,
\end{equation}
%%%%%%%%%%%%%%%%%%%%%%%%%%%%%%%%%%%%%%%%%%%%%%%%%%%%%%%%%%%%%%%%%%%%%%%%%%%%%%%%%%%%%%%%%%%%%%%%%%%%%
where
%%%%%%%%%%%%%%%%%%%%%%%%%%%%%%%%%%%%%%%%%%%%%%%%%%%%%%%%%%%%%%%%%%%%%%%%%%%%%%%%%%%%%%%%%%%%%%%%%%%%%
\[
A_{1} = c^{2}(1+r_{f})(1+r_{0})\, ,\; \; A_{2} = r_{0}r_{f}-1\, ,
\;\; A_{3} = 1+\frac{2(r_{f}-1)}{c^{2}(1+r_{f})}\, ,
\]
\[
A_{4} = [c^{2}(1+r_{f})+2(r_{f}-1)](1+r_{0})\, , \; \; \; A_{5} =
2-\frac{1}{c^{2}(1+r_{f})}.
\]

\begin{figure}[!htb]
  \begin{center}
    \begin{tabular}{ccc}
      \resizebox{110mm}{!}{\includegraphics{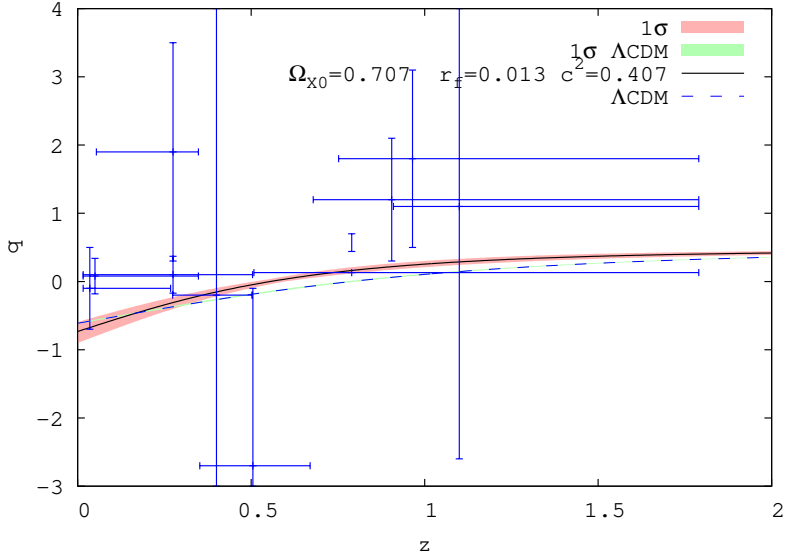}} \\
    \end{tabular}
    \caption{The deceleration parameter vs. redshift for the
    best fit holographic model (solid line) and the $\Lambda$CDM model
    (dashed line). In this, as well as in subsequent figures, the green swath
     indicates the region obtained by including the $1\sigma$ uncertainties of
     the constrained parameters used in the calculation (in the
     present case just $\Omega_{M0}$). The observational data are taken from
     \cite{Daly}; however, in view of the large error bars we do not use these
     data to fit the models.}
    \label{fig:q(z)}
  \end{center}
\end{figure}

Figure \ref{fig:q(z)}  depicts the evolution of the deceleration
parameter, Eq. (\ref{eq:qDeOmega}), for the best fit model. The
observational data are borrowed from  \cite{Daly}. The redshift at
which the universe starts accelerating is $z(q = 0) =
0.56^{+0.12}_{-0.9}$ while for the $\Lambda$CDM model $z(q = 0) =
0.79 \pm 0.02$.
%%%%%%%%%%%%%%%%%%%%%%%%%%%%%%%%%%%%%%%%%%%%%%%%%%%%%%%%%%%%%%%%%%%%%%%%%%%%%%%%%%%%%%%%%%%%%%%%%%%%%
%%%%%%%%%%%%%%%%%%%%%%%%%%%%%%%%%%%%%%%%%%%%%%%%%%%%%%%%%%%%%%%%%%%%%%%%%%%%%%%%%%%%%%%%%%%%%%%%%%%%%
%%%%%%%%%%%%%%%%%%%%%%%%%%%%%%%%%%%%%%%%%%%%%%%%%%%%%%%%%%%%%%%%%%%%%%%%%%%%%%%%%%%%%%%%%%%%%%%%%%%%%

As mentioned above, there is ample freedom in the choice of the
interaction term $Q$. In a previous paper of two of us, on a
holographic dark energy model with the Hubble rate as infrared
cutoff, we took $Q \propto \Omega_{X}$  \cite{id-jcap}. We do not
pursue this possibility here because, as we have checked, it leads
to a universe in which dark energy is subdominant at very late
times. While this does not contradict observation, it looks a bit
odd. In any case, it deserves a separate study which lies beyond
the scope of this paper.

\subsection{Age problem}
Some cosmological models  suffer from the so-called ``age
problem", namely, the existence of high redshifts objects whose
age at some redshift seem to exceed the Universe's age predicted
at that redshift  (as in the $\Lambda$CDM model, see e.g.
\cite{ademir}).

\begin{figure}[!htb]
  \begin{center}
    \begin{tabular}{c}
      \resizebox{120mm}{!}{\includegraphics{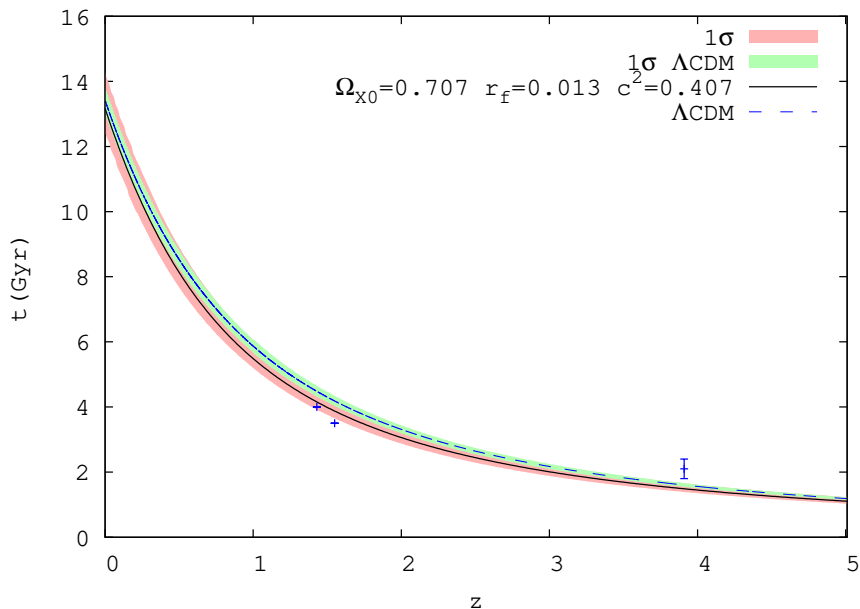}} \\
    \end{tabular}
    \caption{Age of the Universe, $t$, in Giga-years vs. redshift. The solid line  corresponds to
    the best fit holographic model and the dashed line to the $\Lambda$CDM model.
    The data points from left to right locate the old objects LBDS 53W069,
    LBDS 53W091 and APM 08279+5255.}
    \label{fig:t(z)}
  \end{center}
\end{figure}

The age of the Universe at  redshift $z$ is
%%%%%%%%%%%%%%%%%%%%%%%%%%%%%%%%%%%%%%%%%%%%%%%%%%%%%%%%%%%%%%%%%%%%%%%%%%%%%%%%%%%%%%%%
\begin{equation}\label{eq:tDez}
t(z)=t_{0}-\int^{z}_{0}\frac{dz'}{(1+z')\, H(z')} \, .
\end{equation}
%%%%%%%%%%%%%%%%%%%%%%%%%%%%%%%%%%%%%%%%%%%%%%%%%%%%%%%%%%%%%%%%%%%%%%%%%%%%%%%%%%%%%%%%
Figure \ref{fig:t(z)} shows the age of the Universe as a function
of redshift for the best fit holographic model and the
$\Lambda$CDM. Also marked in the figure are the ages and redshifts
of three luminous old objects: galaxies LBDS 53W069 ($z = 1.43$,
$t = 4.0$ Gyr) \cite{dunlop_1999} and LBDS 53W091 ($z = 1.55$, $t
= 3.5$ Gyr) \cite{dunlop_1996,spinrad}, as well as the quasar APM
08279+5255 ($z= 3.91$, $t= 2.1$ Gyr) \cite{hasinger,ademir}. As is
apparent, the ages of two first old objects result compatible with
both, the holographic and the $\Lambda$CDM model; however, the age
of the old quasar falls only within $2\sigma$ with the ages
predicted by these two models. Thus, some tension exists in this
regard.  By contrast, the interacting holographic model of Ref.
\cite{id-jcap}, which takes as infrared cutoff the Hubble radius,
is free of the problem. At any rate, it remains to be seen in
which direction (if any) future observations will ``move" the age
of the said quasar.

%%%%%%%%%%%%%%%%%%%%%%%%%%%%%%%%%%%%Some comment here%%%%%%%%%%%%%%%%%%%%%%%%%%%%%%%%%%%
%%%%%%%%%%%%%%%%%%%%%%%%%%%%%%%%%%%%%%%%%%%%%%%%%%%%%%%%%%%%%%%%%%%%%%%%%%%%%%%%%%%%%%%%
%%%%%%%%%%%%%%%%%%%%%%%%%%%%%%%%%%%%%%%%%%%%%%%%%%%%%%%%%%%%%%%%%%%%%%%%%%%%%%%%%%%%%%%%%%%%%%%%%%%%%
%%%%%%%%%%%%%%%%%%%%%%%%%%%%%%%%%%%%%%%%%%%%%%%%%%%%%%%%%%%%%%%%%%%%%%%%%%%%%%%%%%%%%%%%%%%%%%%%%%%%%
\section{Discussion of the cosmic coincidence}\label{discussion}
In holographic models of dark energy that take the Ricci's length
as the infrared cutoff one can obtain a finite and approximate
constant ratio $r$ for an ample  redshift span even if no
interaction between the dark components is assumed -see e.g.
\cite{gao}. Here we analyze how this comes about, and we note that
while this approach alleviates the coincidence problem it does so
only partially since it would entail that we are living in a
special time.

We start by rewriting  Friedmann's equation (\ref {eq:friedmann1})
with the help of the saturated holographic bound,
$\rho_{X}=3M_{P}^{2}c^{2} \, (\dot{H}+ 2 H^{2})$, as
%%%%%%%%%%%%%%%%%%%%%%%%%%%%%%%%%%%%%%%%%%%%%%%%%%%%%%%%%%%%%%%%%%%%%%%%%%%%%%%%%%%%%%%%%%%%%%%%%%%%%
\begin{equation}\label{eq:Chimento1}
3H^{2}=3c^{2}(\dot{H}+2H^{2})+ \frac{\rho_{M0}}{M_{P}^{2}}\,
(1+z)^{3} \, ,
\end{equation}
%%%%%%%%%%%%%%%%%%%%%%%%%%%%%%%%%%%%%%%%%%%%%%%%%%%%%%%%%%%%%%%%%%%%%%%%%%%%%%%%%%%%%%%%%%%%%%%%%%%%%
and, for convenience, introduce the ancillary variable
$y^{-\alpha}=1+z$. Thus, Eq. (\ref{eq:Chimento1}) takes the form
%%%%%%%%%%%%%%%%%%%%%%%%%%%%%%%%%%%%%%%%%%%%%%%%%%%%%%%%%%%%%%%%%%%%%%%%%%%%%%%%%%%%%%%%%%%%%%%%%%%%%
\begin{equation}\label{eq:Chimento2}
3\alpha^{2}\frac{\dot{y}^{2}}{y^{2}}=
3c^{2}\left\{\alpha\left[\frac{\ddot{y}}{y}-\frac{\dot{y}^{2}}{y^{2}}\right]
+2\alpha^{2}\frac{\dot{y}^{2}}{y^{2}}\right\}+
\frac{\rho_{M0}}{M_{P}^{2}} \, y^{-3\alpha} \,.
\end{equation}
%%%%%%%%%%%%%%%%%%%%%%%%%%%%%%%%%%%%%%%%%%%%%%%%%%%%%%%%%%%%%%%%%%%%%%%%%%%%%%%%%%%%%%%%%%%%%%%%%%%%%
By  equating coefficients, we get $\alpha=c^{2}/(2c^{2}-1)$, and
\begin{equation}
\ddot{y}+\frac{\rho_{M0}}{3c^{2}M_{P}^{2}\,
\alpha}y^{-3\alpha+1}=0 \, . \label{ddotv}
\end{equation}
%%%%%%%%%%%%%%%%%%%%%%%%%%%%%%%%%%%%%%%%%%%%%%%%%%%%%%%%%%%%%%%%%%%%%%%%%%%%%%%%%%%%%%%%%%%%%%%%%%%%%
\noindent Multiplying the latter by $\dot{y}$ the differential
equation can be readily solved. Upon reverting to the original
variable we obtain
%%%%%%%%%%%%%%%%%%%%%%%%%%%%%%%%%%%%%%%%%%%%%%%%%%%%%%%%%%%%%%%%%%%%%%%%%%%%%%%%%%%%%%%%%%%%%%%%%%%%%
\begin{equation}\label{eq:Chimento5}
3H^{2}=\frac{2}{(2-c^{2})M_{P}^{2}}\rho_{M0}(1+z)^{3}+\beta\,
M_{P}^{-2} \rho_{M0}(1+z)^{2\frac{1-2c^{2}}{c^{2}}}\, ,
\end{equation}
%%%%%%%%%%%%%%%%%%%%%%%%%%%%%%%%%%%%%%%%%%%%%%%%%%%%%%%%%%%%%%%%%%%%%%%%%%%%%%%%%%%%%%%%%%%%%%%%%%%%%
\noindent where $\beta$ is a positive-definite integration
constant that can be identified as $\beta = \left(\frac{1}{r_{0}}
\,-\, \frac{c^{2}}{2 \, -\, c^{2}} \right)$ and, of the order of
unity since $r_{0}$ and $c^{2}$ lie not far from $0.4$.
\\\\
Recalling Eq. (\ref{eq:friedmann1}) we finally get
%%%%%%%%%%%%%%%%%%%%%%%%%%%%%%%%%%%%%%%%%%%%%%%%%%%%%%%%%%%%%%%%%%%%%%%%%%%%%%%%%%%%%%%%%%%%%%%%%%%%%
\begin{equation}\label{eq:Chimento6}
\rho_{X}=\frac{c^{2}}{2-c^{2}}\rho_{M0}(1+z)^{3}+\beta\,\rho_{M0}
\, (1+z)^{2\frac{1-2c^{2}}{c^{2}}} \, .
\end{equation}
%%%%%%%%%%%%%%%%%%%%%%%%%%%%%%%%%%%%%%%%%%%%%%%%%%%%%%%%%%%%%%%%%%%%%%%%%%%%%%%%%%%%%%%%%%%%%%%%%%%%%
%%%%%%%%%%%%%%%%%%%%%%%%%%%%%%%%%%%%%%%%%%%%%%%%%%%%%%%%%%%%%%%%%%%%%%%%%%%%%%%%%%%%%%%%%%%%%%%%%%%%%
The DE density is contributed by two terms. The first one
redshifts exactly as non-relativistic matter. The second one, in
view that $c^{2}$ is bounded by $0.36 < c^{2} < 0.8$, results
subdominant for $z$ of order of unity and larger. Therefore, we
can safely conclude that $r = \rho_{M}/\rho_{X} \simeq
(2-c^{2})/c^{2}$ for $0 \leq z $. This is why the ratio $r$
results of order unity in an ample redshift interval, also in the
absence of interaction, as in Ref. \cite{gao}. However, in view of
the observational lower bound on $c^{2}$, we see that $r
\rightarrow 0$ as $z \rightarrow -1$. So, in the holographic
non-interacting model, $r$ results well below unity, close to
zero, and approaches this null value asymptotically for an
infinite span of time. Altogether, according to this model, we
live in a very special and transient period in which $r$ results
comparable to unity.
%%%%%%%%%%%%%%%%%%%%%%%%%%%%%%%%%%%%%%%%%%%%%%%%%%%%%%%%%%%%%%%%%%%%%%%%%%%%%%%%%%%%%%%%%%%%%%%%%%%%%
%%%%%%%%%%%%%%%%%%%%%%%%%%%%%%%%%%%%%%%%%%%%%%%%%%%%%%%%%%%%%%%%%%%%%%%%%%%%%%%%%%%%%%%%%%%%%%%%%%%%%
%%%%%%%%%%%%%%%%%%%%%%%%%%%%%%%%%%%%%%%%%%%%%%%%%%%%%%%%%%%%%%%%%%%%%%%%%%%%%%%%%%%%%%%%%%%%%%%%%%%%%

\section {Observational constraints}\label{statistical-analysis}
To constrain the four free parameters ($\Omega_{X0}$, $c^{2}$,
$r_{f}$, and $H_{0}$) of the holographic model presented above we
use observational data from SN Ia (557 data points), the
CMB-shift, BAO, and gas mass fractions in galaxy clusters as
inferred from x-ray data (42 data points), the Hubble rate (15
data points), and the growth function (5 data points). Being the
likelihood function defined as $ {\cal L} \propto \exp (-
\chi^{2}/2)$ the best fit follows  from minimizing the sum $
\chi^{2}_{\rm total} = \chi^{2}_{sn} \, + \, \chi^{2}_{cmb}\, + \,
\chi^{2}_{bao} \, + \, \chi^{2}_{x-rays}+ \, \chi^{2}_{Hubble}\, +
\, \chi^{2}_{gf}$.
\\\\
\subsection {SN Ia}
We contrast the theoretical distance modulus
%%%%%%%%%%%%%%%%%%%%%%%%%%%%%%%%%%%%%%%%%%%%%%%%%%%%%%%%%%%%%%%%%%%%%
\be \mu_{th}(z_{i}) = 5\log_{10}\left(\frac{D_{L}}{{10{\rm
pc}}}\right)\, + \,\mu_{0}\, , \label{modulus} \ee
%%%%%%%%%%%%%%%%%%%%%%%%%%%%%%%%%%%%%%%%%%%%%%%%%%%%%%%%%%%%%%%%%%%
where $\mu_{0} = 42.38 \, -\, 5\log_{10} h$, with the observed
distance modulus $\mu_{obs}(z_{i})$ of the 557 SN Ia compiled in
the Union2 set \cite{Amanullah}. The latter assemble is much
richer than previous SN Ia compilations and has some other
advantages, especially the refitting of all light curves with the
SALT2 fitter and an enhanced control of systematic errors. In
(\ref{modulus}) $D_{L} = (1+z) \int_{0}^{z}{\frac{dz'}{E(z';{\bf
p})}}$ denotes the Hubble-free luminosity distance, with ${\bf p}$
the model parameters ($ \Omega_{X0}$, $c^{2}$, $r_{f}$, and $H_{0}
$), and $E(z; {\bf p}) := H(z; {\bf p})/H_{0}$.

The $\chi^{2}$ from the 557 SN Ia is given by
%%%%%%%%%%%%%%%%%%%%%%%%%%%%%%%%%%%%%%%%%%%%%%%%%%%%%%%%%%%%%%%%%%
\be \chi^{2}_{sn}({\bf p}) = \sum_{i=1}^{557}
\frac{[\mu_{th}(z_{i}) \, - \,
\mu_{obs}(z_{i})]^{2}}{\sigma^{2}(z_{i})} \, , \label{chi2mu} \ee
%%%%%%%%%%%%%%%%%%%%%%%%%%%%%%%%%%%%%%%%%%%%%%%%%%%%%%%%%%%%%%%%%%%
where $\sigma_{i}$ denotes the 1$\sigma$ uncertainty associated to
the $i$th data point.

To eliminate the effect of the nuisance parameter $\mu_{0}$ we
resort to the method of \cite{Nesseris-Perilovorapoulos_0} to
obtain $\tilde{\chi}^{2}_{sn} = \chi^{2\, (minimum)}_{sn} =
543.70$.
%%%%%%%%%%%%%%%%%%%%%%%%%%%%%%%%%%%%%%%%%%%%%%%%%%%%%%%%%%%%%%%%%%%%%%%%%%%%%%%%%%%%%%
%
\subsection{CMB-shift}
The displacement of the first acoustic peak of the CMB temperature
spectrum with respect to the location it would take should the
Universe be described by the Einstein-de Sitter model is given by
the CMB-shift \cite{wang-mukherjee,bond}
%%%%%%%%%%%%%%%%%%%%%%%%%%%%%%%%%%%%%%%%%%%%%%%%%%%%%%%%%%%%%%%%%%%%%%%%%%%%%%%%%%%%%%
\begin{equation}\label{eq:CMBShift}
{\cal R} =\sqrt{\Omega_{M_{0}}}\int^{z_{rec}}_{0}\frac{dz}{E(z;
{\bf p)}} \, ,
\end{equation}
%%%%%%%%%%%%%%%%%%%%%%%%%%%%%%%%%%%%%%%%%%%%%%%%%%%%%%%%%%%%%%%%%%%%%%%%%%%%%%%%%%%%%%
where $z_{rec}\simeq 1089$ is the redshift at the recombination
epoch. This parameter is approximately model-independent but not
quite as the above expression somehow assumes the $\Lambda$CDM
model.

The 7-year WMAP data provides ${\cal R}(z_{rec})=1.725 \pm 0.018$
\cite{komatsu}. The best fit value of the model is ${\cal
R}(z_{rec})=1.727 \pm 0.030$. Minimization of
%%%%%%%%%%%%%%%%%%%%%%%%%%%%%%%%%%%%%%%%%%%%%%%%%%%%%%%%%%%%%%%%%%%%%%%%%%
\be \chi^{2}_{cmb} ({\bf p})= \frac{({\cal R}_{th} \, - \, {\cal
R}_{obs})^{2}}{\sigma^{2}_{{\cal R}}} \, \label{chi2R} \ee
%%%%%%%%%%%%%%%%%%%%%%%%%%%%%%%%%%%%%%%%%%%%%%%%%%%%%%%%%%%%%%%%%%%%%%%%%%
produces $\chi^{2\, (minimum)}_{CMB-shift} = 0.013$.
\\\\
\subsection{BAO}
Pressure waves originated from cosmological perturbations in the
primeval baryon-photon plasma produced acoustic oscillations in
the baryonic fluid. These oscillations have been unveiled by a
clear peak in the large scale correlation function measured from
the luminous red galaxies sample of the Sloan Digital Sky Survey
(SDSS)  at $z = 0.35$ \cite{Eisenstein} as well as in the Two
Degree Field Galaxy Redshift Survey (2dFGRS) at $z = 0.2$
\cite{Percival}. These peaks can be traced to expanding spherical
waves of baryonic perturbations with a characteristic distance
scale
%%%%%%%%%%%%%%%%%%%%%%%%%%%%%%%%%%%%%%%%%%%%%%%%%%%%%%%%%%%%%%%%%%%%%%%%%%%%
\begin{equation}\label{eq:BAO}
D_{v}(z_{BAO})=\left[\frac{z_{BAO}}{H(z_{BAO})}
\left(\int^{z_{BAO}}_{0}\frac{dz}{H(z)}\right)^{2}\right]^{\frac{1}{3}}
\,
\end{equation}
%%%%%%%%%%%%%%%%%%%%%%%%%%%%%%%%%%%%%%%%%%%%%%%%%%%%%%%%%%%%%%%%%%%%%%%%%%%%%
-see e.g. \cite{Nesseris-Perilovorapoulos-jcap}.

Data from SDSS  and 2dFGRS observations yield
$D_{v}(0.35)/D_{v}(0.2)=1.736 \pm 0.065 \, $ \cite{Percival}. The
best fit value for the holographic model is
$D_{v}(0.35)/D_{v}(0.2) = 1.664 \pm 0.003$, and minimization of
%%%%%%%%%%%%%%%%%%%%%%%%%%%%%%%%%%%%%%%%%%%%%%%%%%%%%%%%%%%%%%%%%%%%%%%%%%%%
\be \chi^{2}_{bao} ({\bf p}) =
\frac{([D_{v}(0.35)/D_{v}(0.2)]_{th} \, - \,
[D_{v}(0.35)/D_{v}(0.2)]_{obs})^{2}}{\sigma^{2}_{D_{v}(0.35)/D_{v}(0.2)}}
\label{chi2BAO} \ee
%%%%%%%%%%%%%%%%%%%%%%%%%%%%%%%%%%%%%%%%%%%%%%%%%%%%%%%%%%%%%%%%%%%%%%%%%%%%
gives $\chi^{2\, (minimum)}_{bao} = 1.20$.
\\\\
\subsection {Gas mass fraction}
As is well known, a very useful indicator of the overall cosmic
ratio $\Omega_{baryons}/\Omega_{M}$, nearly independent of
redshift,  is the fraction of baryons in galaxy clusters
\cite{White}. This quantity can be determined from the x-ray flux
originated in hot clouds of baryons and it is related to the
cosmological parameters by $f_{gas} \propto d_{A}^{3/2}$, where
$d_{A}: = (1+z)^{-1}\int_{0}^{z}{\frac{dz'}{H(z')}}$ denotes the
angular diameter distance to the cluster.

We used measurements by the Chandra satellite of 42 dynamically
relaxed galaxy clusters in the redshift interval $0.05 < z < 1.1$
\cite{Allen}. In  fitting the data we resorted to the empirical
formula
%%%%%%%%%%%%%%%%%%%%%%%%%%%%%%%%%%%%%%%%%%%%%%%%%%%%%%%%%%%%%%
\begin{equation}\label{eq:Allen}
f_{gas}(z)=\frac{K \, A \, \gamma\, b(z)}{1\, + \, s(z)}
\frac{\Omega_{B0}}{\Omega_{M0}}\left(\frac{d_{A}^{\Lambda
CDM}}{d_{A}}\right)^{3/2}
\end{equation}
%%%%%%%%%%%%%%%%%%%%%%%%%%%%%%%%%%%%%%%%%%%%%%%%%%%%%%%%%%%%%%
(see Eq. (3) in Ref. \cite{Allen}) in which the $\Lambda$CDM model
served  as a reference. Here, the parameters $K$, $A$, $\gamma$,
$b(z)$ and $s(z)$ model the aboundance of gas in the clusters. We
set these parameters to their respective best fit values of Ref.
\cite{Allen}.

The $\chi^{2}$ function from the 42 galaxy clusters reads
%%%%%%%%%%%%%%%%%%%%%%%%%%%%%%%%%%%%%%%%%%%%%%%%%%%%%%%%%%%%%%%%%%
\be
\chi^{2}_{x-rays}({\bf p}) = \sum_{i=1}^{42}
\frac{([f_{gas}(z_{i})]_{th} \, - \,
[f_{gas}(z_{i})]_{obs})^{2}}{\sigma^{2}(z_{i})} \, ,
\label{chi2fgas}
\ee
%%%%%%%%%%%%%%%%%%%%%%%%%%%%%%%%%%%%%%%%%%%%%%%%%%%%%%%%%%%%%%%%%%%
and its minimum value comes to be $\chi^{2\, (minimum)}_{x-rays} =
41.79$.

\begin{figure}[!htb]
  \begin{center}
    \begin{tabular}{ccc}
      \resizebox{110mm}{!}{\includegraphics{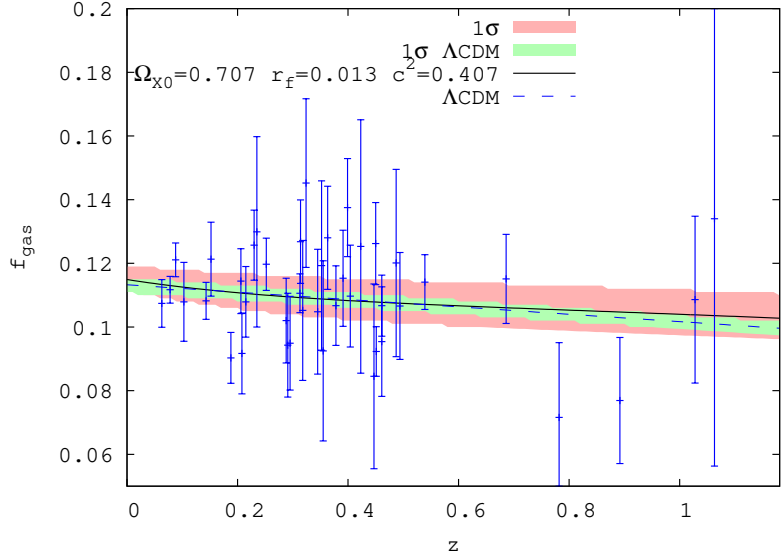}} \\
    \end{tabular}
    \caption{Gas mass fraction in 42 relaxed galaxy clusters vs.
    redshift. The solid and dashed curves correspond to the best fit
    models: holographic and $\Lambda$CDM, respectively.
    The data points with their error bars are taken from Table III
    of Ref. \cite{Allen}.}
    \label{fig:XRay}
  \end{center}
\end{figure}

%%%%%%%%%%%%%%%%%%%%%%%%%%%%%%%%%%%%%%%%%%%%%%%%%%%%%%%%%%%%%%%%%%%%%%%%
Figure \ref{fig:XRay} shows the fit to the data.

%%%%%%%%%%%%%%%%%%%%%%%%%%%%%%%%%%%%%%%%%%%%%%%%%%%%%%%%%%%%%%%%%%%%%%%%%%%%%
%%%%%%%%%%%%%%%%%%%%%%%%%%%%%%%%%%%%%%%%%%%%%%%%%%%%%%%%%%%%%%%%%%%%%%%%%%%%%
\subsection {History of the Hubble parameter}
Recently, high precision measurements by Riess {\em et al.} at $z
= 0$, from the observation of 240 Cepheid variables of rather
similar periods and metallicities \cite{Riess-2009}, as well as
measurements by Gazta\~{n}aga {\em et al.}, at $z = 0.24, \, 0.34,
{\rm and}\, 0.43 \,$ \cite{gazta}, who used the BAO peak position
as a standard ruler in the radial direction, have somewhat
improved our knowledge about $H(z)$. However, at redshifts above,
say, $0.5$ this function remains largely unconstrained. Yet, in
order to constrain the holographic model we have employed these
four data alongside 11 noisier data in the redshift interval $0.1
\lesssim z \lesssim 1.8$, from Simon {\em et al.} \cite{simon} and
Stern {\em et al.} \cite{stern}, obtained from the differential
ages of passive-evolving galaxies and archival data.

\begin{figure}[!htb]
  \begin{center}
    \begin{tabular}{ccc}
      \resizebox{110mm}{!}{\includegraphics{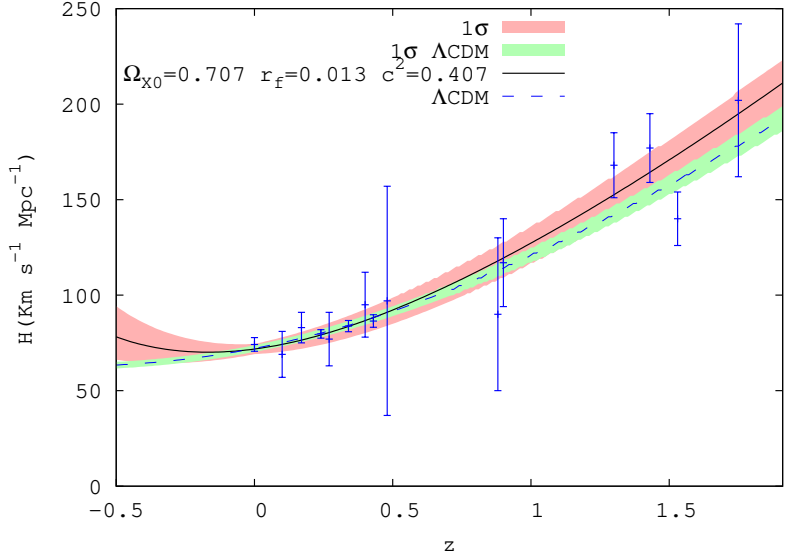}} \\
    \end{tabular}
    \caption{History of the Hubble factor in terms of the redshift for
     the best fit values of the holographic model (solid line) and the
     $\Lambda$CDM model (dashed line). The data points and error bars
     are borrowed from Refs. \cite{Riess-2009,gazta,simon,stern}.}
    \label{fig:H(z)}
  \end{center}
\end{figure}

By minimizing
%%%%%%%%%%%%%%%%%%%%%%%%%%%%%%%%%%%%%%%%%%%%%%%%%%%%%%%%%%%%%%%%%%%%%%%%%
\be \chi^{2}_{Hubble}({\bf p}) = \sum_{i=1}^{15}
\frac{[H_{th}(z_{i}) \, - \,
H_{obs}(z_{i})]^{2}}{\sigma^{2}(z_{i})}
 \label{chi2hubble}
 \ee
%%%%%%%%%%%%%%%%%%%%%%%%%%%%%%%%%%%%%%%%%%%%%%%%%%%%%%%%%%%%%%%%%%%%%%%%%
we got $\chi^{2\, (minimum)}_{Hubble} = 9.57$ and $H_{0} = 71.8
\pm 2.9 \,$ km/s/Mpc as the best fit for the Hubble's constant.
Figure \ref{fig:H(z)} depicts the Hubble history according to the
best fit holographic model and the best $\Lambda$CDM model.
%%%%%%%%%%%%%%%%%%%%%%%%%%%%%%%%%%%%%%%%%%%%%%%%%%%%%%%%%%%%%%%%%%%%%%%%%
%%%%%%%%%%%%%%%%%%%%%%%%%%%%%%%%%%%%%%%%%%%%%%%%%%%%%%%%%%%%%%%%%%%%%%%%%%%%%
%%%%%%%%%%%%%%%%%%%%%%%%%%%%%%%%%%%%%%%%%%%%%%%%%%%%%%%%%%%%%%%%%%%%%%%%%%%%%

\subsection {Growth function}
Up to now, we have considered background quantities that chiefly
depend on $H(z)$ whence they are not very useful at discriminating
between cosmological models that present a similar Hubble history,
independently of how different they are otherwise. One manner to
circumvent this hurdle is to study evolution of the growth
function $f = d\ln\delta_{M}/d\ln a$, where $\delta_{M}$ denotes
the density contrast of non-relativistic matter.

\begin{figure}[!htb]
  \begin{center}
    \begin{tabular}{ccc}
      \resizebox{60mm}{!}{\includegraphics{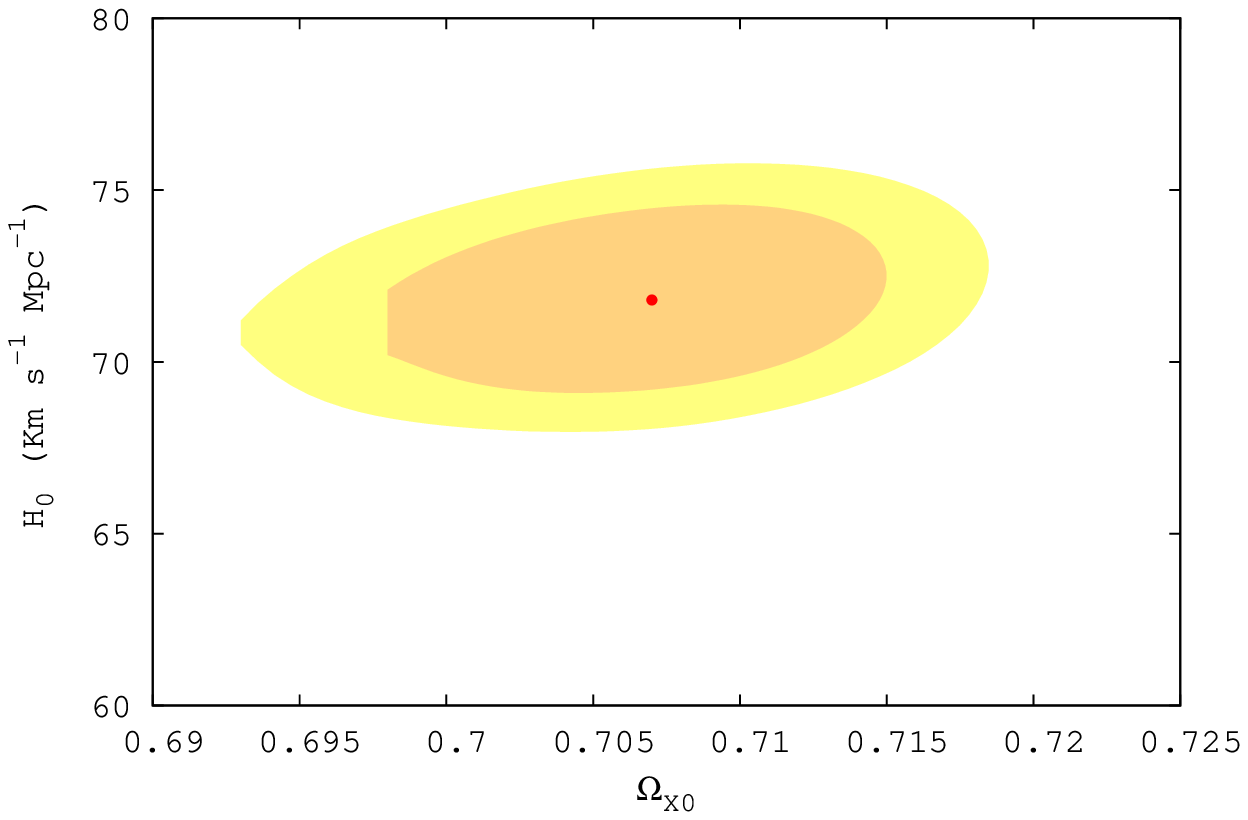}} &
      \resizebox{60mm}{!}{\includegraphics{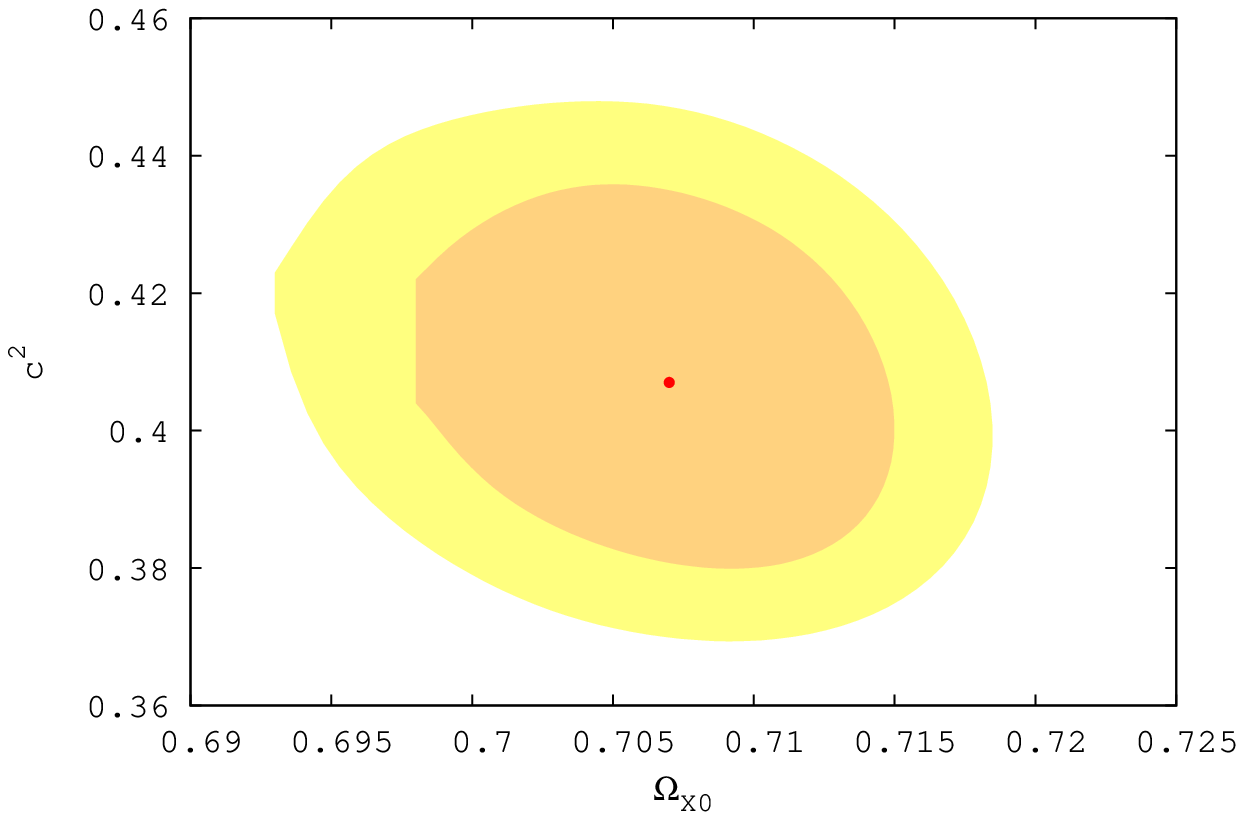}} &
      \resizebox{60mm}{!}{\includegraphics{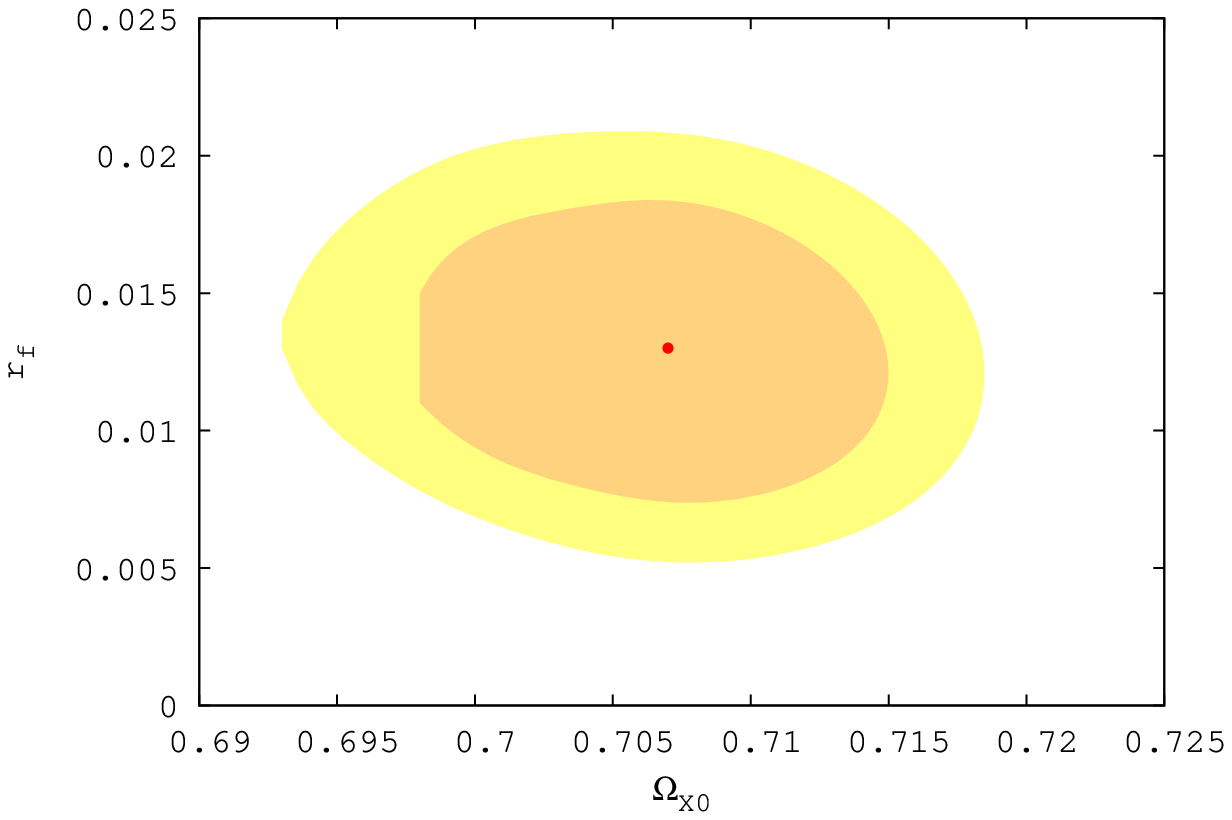}}\\
    \end{tabular}
    \caption{Panels from left to right show the 68.3\% and 95.4\% confidence
    contours for the pair of free parameters  ($\Omega_{X0}$, $H_{0}$),
    ($\Omega_{X0}$, $c^{2}$), ($\Omega_{X0}$, $r_{f}$), respectively,
     obtained by constraining the holographic model with
     SN Ia+CMB-shift+ BAO+x-ray+H(z)+growth function data.
     The solid point in each panel locates the best fit values.}
    \label{fig:ellipses1}
  \end{center}
\end{figure}

\begin{figure}[!htb]
  \begin{center}
    \begin{tabular}{ccc}
      \resizebox{60mm}{!}{\includegraphics{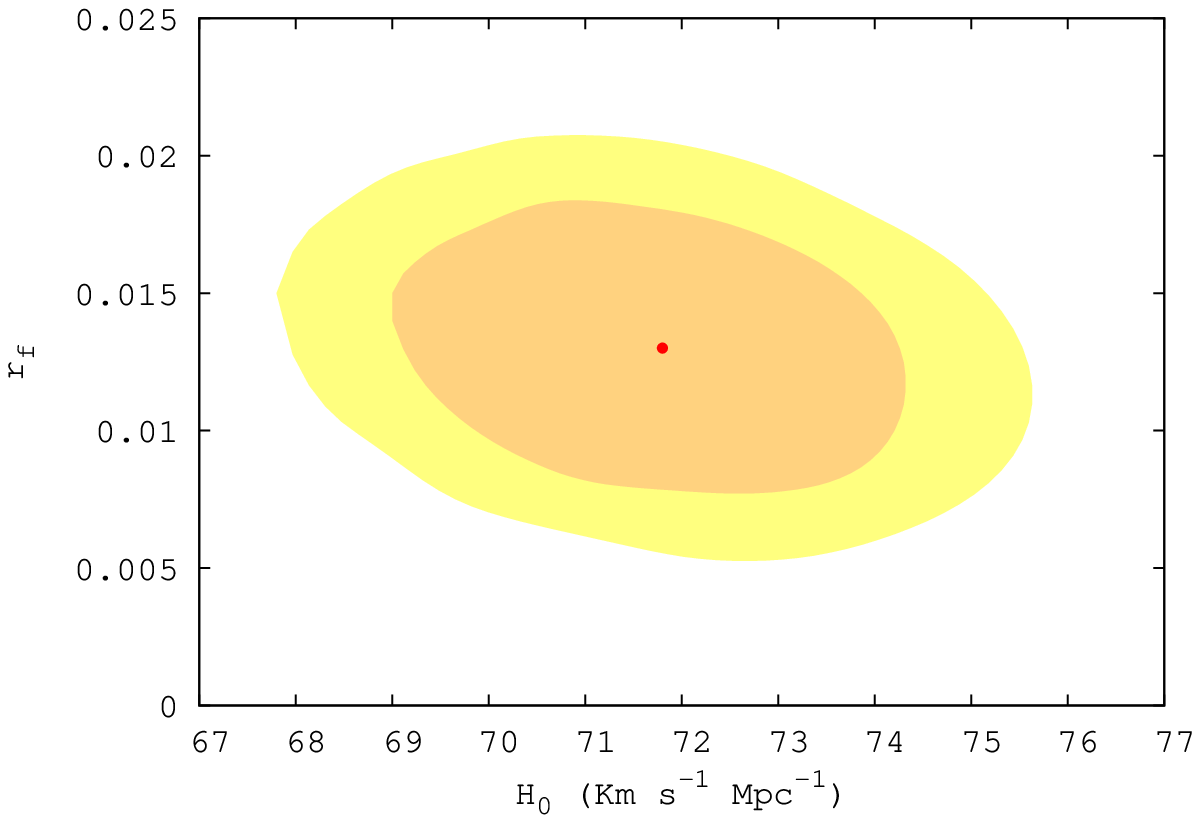}} &
      \resizebox{60mm}{!}{\includegraphics{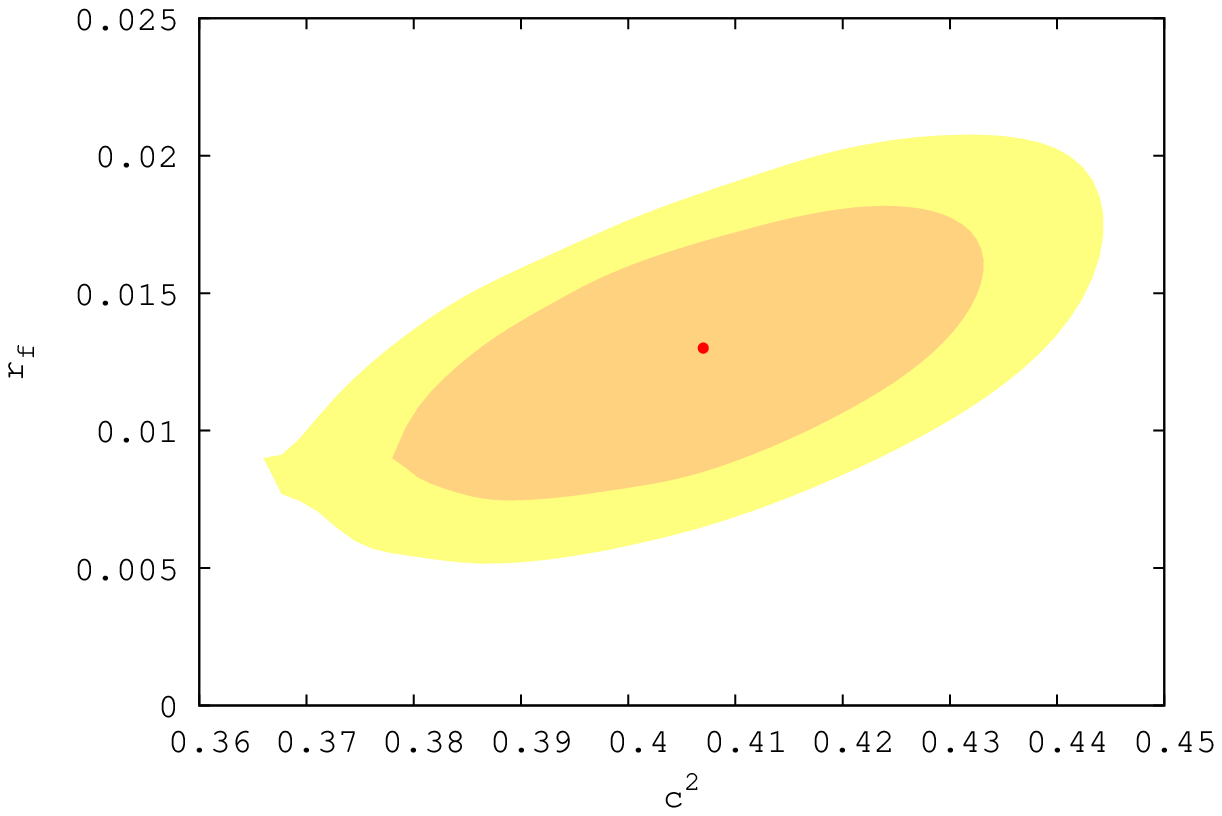}} &
      \resizebox{60mm}{!}{\includegraphics{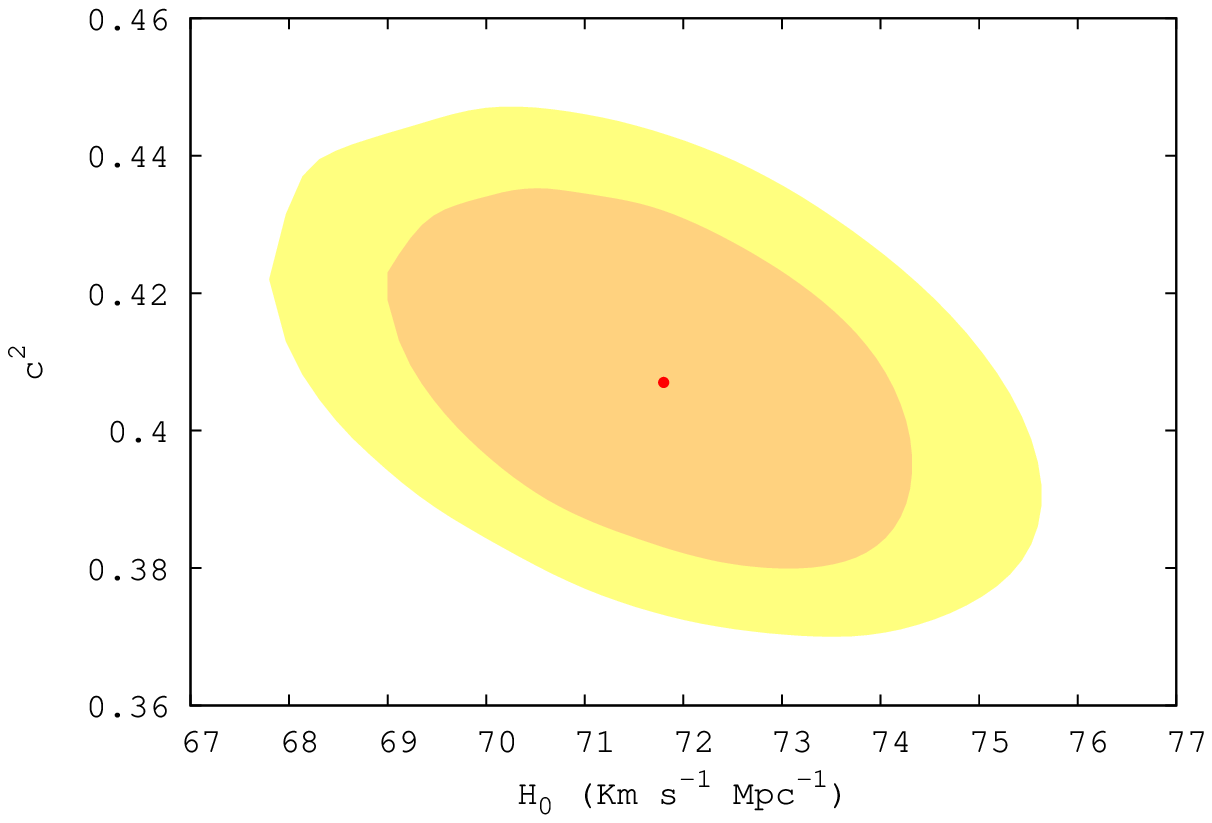}}\\
    \end{tabular}
    \caption{Same as Fig. \ref{fig:ellipses1} for the pairs of free
     parameters  ($H_{0}$, $r_{f}$), ($c^{2}$, $r_{f}$), and ($H_{0}$, $c^{2}$).}
    \label{fig:ellipses2}
  \end{center}
\end{figure}

\begin{figure}[!htb]
  \begin{center}
    \begin{tabular}{ccc}
      \resizebox{90mm}{!}{\includegraphics{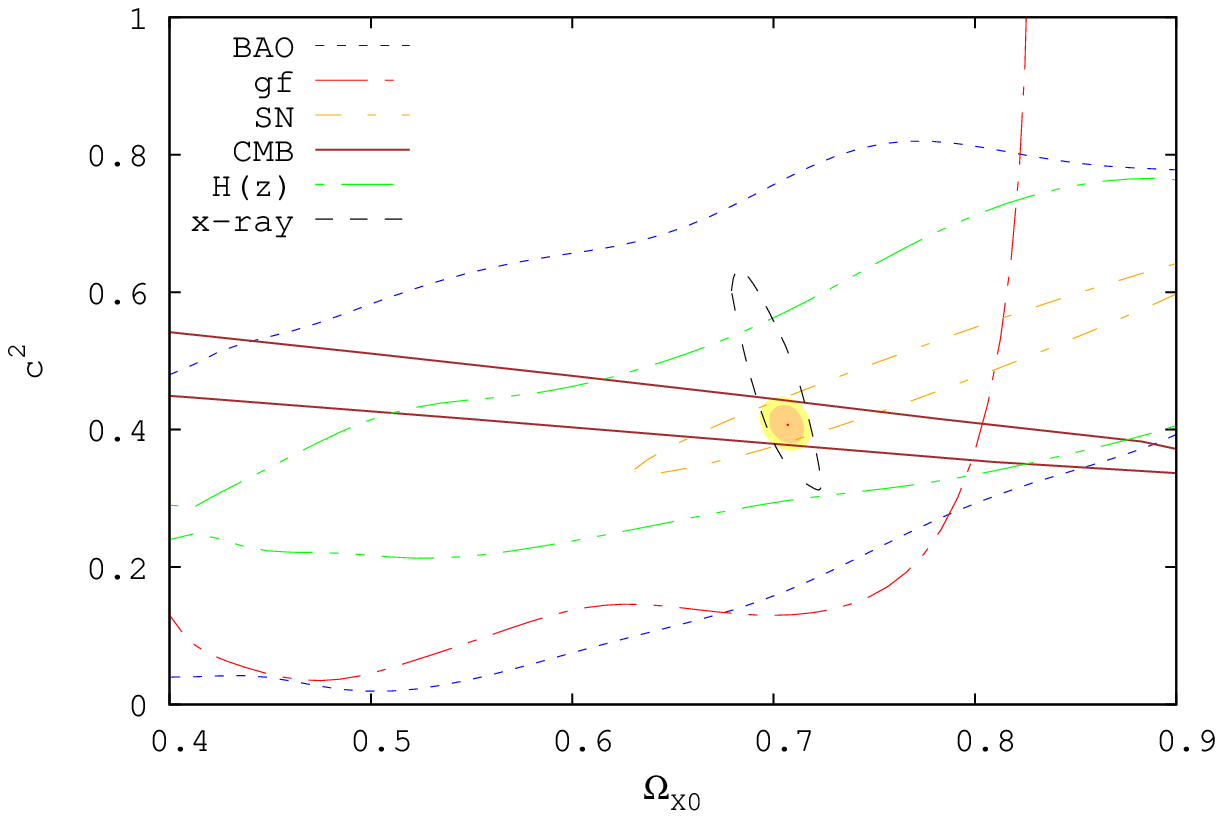}} &
      \resizebox{90mm}{!}{\includegraphics{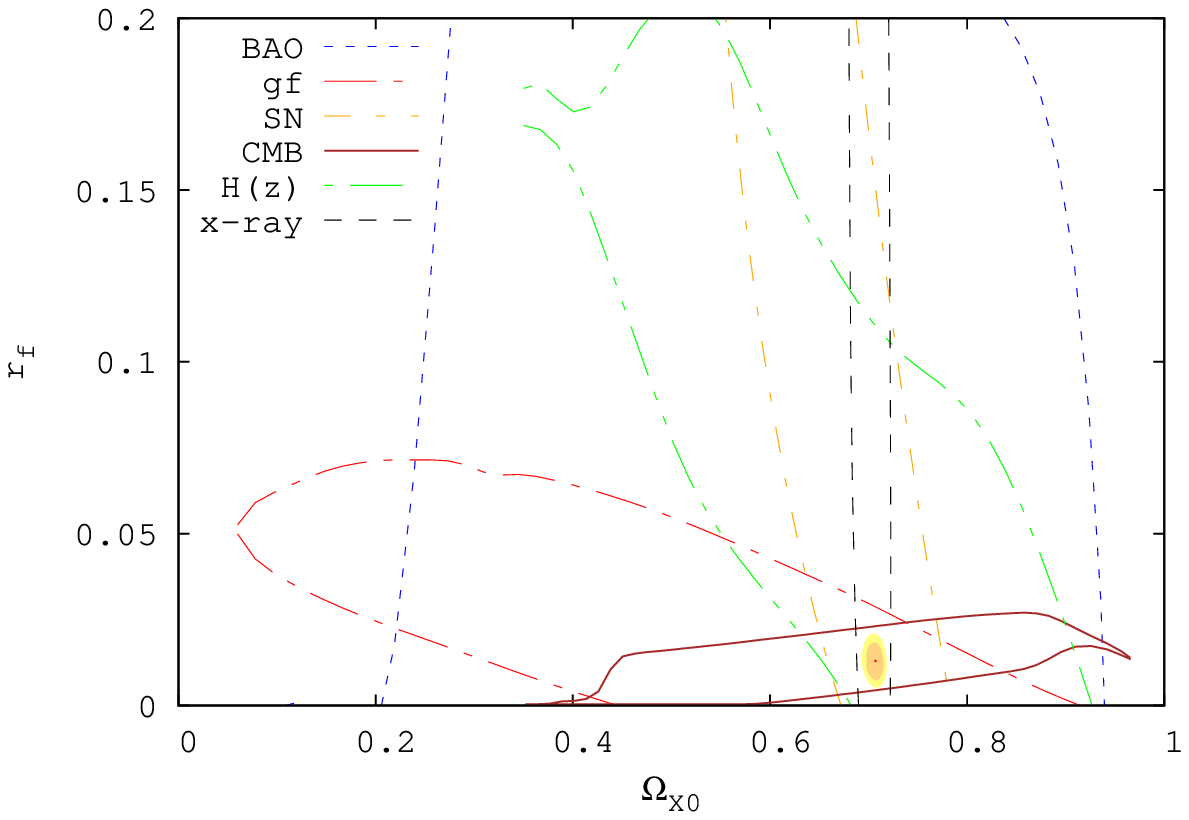}}\\
    \end{tabular}
    \caption{Left panel: The probability contours for SN Ia, CMB, BAO, x-ray, $H(z)$, and growth function,
    in the ($\Omega_{X0}, c^{2}$) plane. The joined constraint uses $ \chi^{2}_{\rm total} =
    \chi^{2}_{sn} \, + \, \chi^{2}_{cmb}\, + \, \chi^{2}_{bao} \, + \, \chi^{2}_{x-rays}+ \,
    \chi^{2}_{Hubble}\, + \, \chi^{2}_{gf}$. Right panel: Idem
    for the ($\Omega_{X0}, r_{f}$) plane.}
    \label{fig:ellipses3}
  \end{center}
\end{figure}

The evolution of the latter obeys the coupled set of equations
%%%%%%%%%%%%%%%%%%%%%%%%%%%%%%%%%%%%%%%%%%%%%%%%%%%%%%%%%%%%%%%%%%%%%%%%%%%%%%%%%%%%%%%%
\begin{eqnarray}\label{eq:densityContr1}
\dot{\delta}_{M}-\frac{k^{2}}{a}v_{M}=-\frac{1+r}{r}Q \,H \,\delta_{M} \, ,\\
\label{eq:densityContr2} \dot{v}_{M}+Hv_{M}+\frac{1}{a}\phi=0 \, ,
\end{eqnarray}
%%%%%%%%%%%%%%%%%%%%%%%%%%%%%%%%%%%%%%%%%%%%%%%%%%%%%%%%%%%%%%%%%%%%%%%%%%%%%%%%%%%%%%%%
where the Newtonian potential $\phi$ fulfills Poisson's equation
%%%%%%%%%%%%%%%%%%%%%%%%%%%%%%%%%%%%%%%%%%%%%%%%%%%%%%%%%%%%%%%%%%%%%%%%%%%%%%%%%%%%%%%%
\begin{equation}\label{eq:Poisson}
\frac{k^{2}}{a^{2}}=-4\pi G \rho_{M}\delta_{M} \, .
\end{equation}
%%%%%%%%%%%%%%%%%%%%%%%%%%%%%%%%%%%%%%%%%%%%%%%%%%%%%%%%%%%%%%%%%%%%%%%%%%%%%%%%%%%%%%%%
%%%%%%%%%%%%%%%%%%%%%%%%%%%%%%%%%%%%%%%%%%%%%%%%%%%%%%%%%%%%%%%%%%%%%%%%%%%%%%%%%%%%%%%%
%%%%%%%%%%%%%%%%%%%%%%%%%%%%%%%%%%%%%%%%%%%%%%%%%%%%%%%%%%%%%%%%%%%%%%%%%%%%%%%%%%%%%%%%
Solving the equations, and expressing in terms of $r$ and using
$w=[c^{2}(1\, + \, r)\, -\, 2]/(3c^{2})$ obtained from
(\ref{eq:Omega}), $c^{-2}(1\, +\, r)^{-1} = (\dot{H}/H^{2}) \, +
\, 2$, and (\ref{eq:rDotChim}), we get from the evolution
equations (\ref{eq:EvolIn}) and (\ref{eq:EvolIn2})
%%%%%%%%%%%%%%%%%%%%%%%%%%%%%%%%%%%%%%%%%%%%%%%%%%%%%%%%%%%%%%%%%%%%%%%%%%%%%%%%%%%%%%%%
\begin{equation}\label{eq:growthFactor}
f'+f^{2}+\left(\frac{1}{c^{2}(1+r)}+Q \frac{1+r}{r}\right)f-
\frac{3r^{3}+2Q^{2}(1+r)^{3}-2Q r(2+r-r^{2})}{2r^{2}(1+r)}=0 \, ,
\end{equation}
%%%%%%%%%%%%%%%%%%%%%%%%%%%%%%%%%%%%%%%%%%%%%%%%%%%%%%%%%%%%%%%%%%%%%%%%%%%%%%%%%%%%%%%%
where $f' \equiv df/d\ln a$. Note that in the limit $Q \rightarrow
0$, last equation collapses to the corresponding expression of the
Einstein-de Sitter scenario ($\Omega_{M}(z) = 1$ and $\delta_{M}
\propto a \propto t^{2/3}$); that is to say, $f' + f^{2} + [2 \, +
\, (\dot{H}/H^{2})] f = 3\Omega_{M}/2$. (Recall that in
Einstein-de Sitter  $\dot{H}/H^{2} = -3/2$ and the solution of the
equation for $f$ is simply $f = 1$).
%%%%%%%%%%%%%%%%%%%%%%%%%%%%%%%%%%%%%%%%%%%%%%%%%%%%%%%%%%%%%%%%%%%%%%%%%%%%%%%%%%%%%%%%
\begin{figure}[!htb]
  \begin{center}
    \begin{tabular}{ccc}
      \resizebox{110mm}{!}{\includegraphics{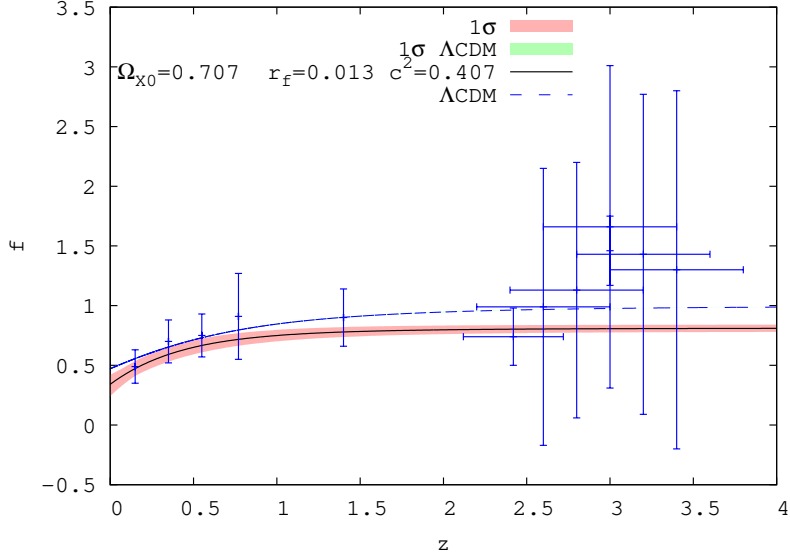}} \\
    \end{tabular}
    \caption{Growth function, $f$, vs. redshift as follows from integrating
    Eq. (\ref{eq:growthFactor})(solid line).
    Also shown is the prediction of the $\Lambda$CDM model (dashed line).
    The observational data were taken from Ref. \cite{Gong}. In constraining
    both models we have used only the five less noisy data depicted in the
    figure, (i.e., data corresponding to redshifts below $1.5$).}
    \label{fig:gf}
  \end{center}
\end{figure}
%%%%%%%%%%%%%%%%%%%%%%%%%%%%%%%%%%%%%%%%%%%%%%%%%%%%%%%%%%%%%%%%%%%%%%%%%%%%%%%%%%%%%%%%%%%%%%%%%%%%%
%%%%%%%%%%%%%%%%%%%%%%%%%%%%%%%%%%%%%%%%%%%%%%%%%%%%%%%%%%%%%%%%%%%%%%%%%%%%%%%%%%%%%%%%%%%%%%%%%%%%%
%%%%%%%%%%%%%%%%%%%%%%%%%%%%%%%%%%%%%%%%%%%%%%%%%%%%%%%%%%%%%%%%%%%%%%%%%%%%%%%%%%%%%%%%%%%%%%%%%%%%%
In constraining the model we have taken only the five lowest
redshift data of the growth function shown in Fig. \ref{fig:gf}
-the other data present very large error bars. The best fit yields
$\chi^{2} = 1.06$.

\[\]
Figures \ref{fig:ellipses1} - \ref{fig:ellipses3} and table I
summarize our findings.

%%%%%%%%%%%%%%%%%%%%%%%%%%%%%%%%%%%%%%%%%%%%%%%%%%%%%%%%%%%%%%%%%%%%%%%%%%%%%%%%%%%%%%%%

Figure \ref{fig:ellipses3} depicts the $1\sigma$ confidence
contours from SN Ia (dashed yellow), CMB-shift (solid black), BAO
(dashed blue), x-rays (dashed black), history of the Hubble
function (dot-dot dashed green), and grow function (dot-dashed
red) in the ($\Omega_{X0}, c^{2}$) plane (left panel) and the
($\Omega_{X0}, r_{f}$) plane. The joined constraints corresponding
to $\chi^{2}_{total}$ are shown as shaded contours. As is apparent
from left panel most of the discriminatory power arises from the
near orthogonality between the  x-ray and CMB-shift and supernovae
contours. However, in the right panel the supernovae contour
appears nearly degenerated with respect to the x-ray contour.

%%%%%%%%%%%%%%%%%%%%%%%%%%%%%%%%%%%%%%%%%%%%%%%%%%%%%%%%%%%%%%%%%%%%%%%%%%%%%%%%%%%%%%%%%
%%%%%%%%%%%%%%%%%%%%%%%%%%%%%%%%%%%%%%%%%%%%%%%%%%%%%%%%%%%%%%%%%%%%%%%%%%%%%%%%%%%%%%%%%
%%%%%%%%%%%%%%%%%%%%%%%%%%%%%%%%%%%%%%%%%%%%%%%%%%%%%%%%%%%%%%%%%%%%%%%%%%%%%%%%%%%%%%%%%%%%%%%%%%%
%%%%%%%%%%%%%%%%%%%%%%%%%%%%%%%%%%%%%%%%%%%%%%%%%%%%%%%%%%%%%%%%%%%%%%%%%%%%%%%%%%%%%%%%%%%%%%%%%%%
%%%%%%%%%%%%%%%%%%%%%%%%%%%%%%%%%%%%%%%%%%%%%%%%%%%%%%%%%%%%%%%%%%%%%%%%%%%%%%%%%%%%%%%%%%%%%%%%%%%%%
%%%%%%%%%%%%%%%%%%%%%%%%%%%%%%%%%%%%%%%%%%%%%%%%%%%%%%%%%%%%%%%%%%%%%%%%%%%%%%%%%%%%%%%%%%%%%%%%%%%%%
%%%%%%%%%%%%%%%%%%%%%%%%%%%%%%%%%%%%%%%%%%%%%%%%%%%%%%%%%%%%%%%%%%%%%%%%%%%%%%%%%%%%%%%%%%%%%%%%%%%%%
%%%%%%%%%%%%%%%%%%%%%%%%%%%%%%%%%%%%%%%%%%%%%%%%%%%%%%%%%%%%%%%%%%%%%%%%%%%%%%%%%%%%%%%%%%%%%%%%%%%%%

Altogether, by constraining the holographic model of Section II
with data from SN Ia, CMB-shif, BAO, x-rays, H(z), and the growth
function we obtain $\Omega_{X0}=0.707\pm0.009$,
$c^{2}=0.407^{+0.033}_{-0.028}$, $r_{f}= 0.013^{+0.006}_{-0.005}$,
and $H_{0}=71.8\pm2.9$ km/s/Mpc as best fit parameters. It is
worth noticing that the non-interacting case, $Q = 0$ (which
implies $r_{f} = 0$ via Eq. (\ref{eq:QChim})), lies over $2\sigma$
away from the best fit value. This feature seems typical of
holographic dark energy models (see e.g. \cite{gao, suwa,
id-jcap}).

Table \ref{table:chi2} shows the partial, total, and total
$\chi^{2}$ over the number of degrees of freedom of the
holographic model along with the corresponding values for the
$\Lambda$CDM model. In the latter one has just two free
parameters, $\Omega_{M0}$ and $H_{0}$. Their best fit values after
constraining the model to the data are $\Omega_{M0} = 0.266 \pm
0.006$, and $H_{0} = 71.8 \pm 1.9 \, $km/s/Mpc.

%%%%%%%%%%%%%%%%%%%%%%%%%%%%%%%%%%%%%%%%%%%%%%%%%%%%%%%%%%%%%%%%%%%%%%%%%%%%%%%%%%%%%%%%%%%%%%%%%%%%%%%%%%%%%%%
\begin{table}
\begin{tabular}{ p{2.4 cm}| p{1.80 cm} p{1.4 cm} p{1.4 cm} p{1.6 cm}p{1.6 cm}p{1.6 cm} p{1.8 cm} p{1.4
cm}}
\hline \hline
Model & $\chi^{2}_{sn}$ & $\chi^{2}_{cmb}$ & $\chi^{2}_{bao}$ &  $\chi^{2}_{x-rays}$ &  $\chi^{2}_{H}$ & $\chi^{2}_{gf}$ & $\chi^{2}_{{\rm total}}$ & $\chi^{2}_{{\rm total}}/dof$  \\
\hline
Holographic  & $543.70$ & $0.01$ & $1.20$ & $41.79$ & $9.57$ & 1.06 & $597.34$ & $\; \; 0.96$ \\
\hline
$\Lambda$CDM & $542.87$ & $0.05$ & $1.13$ & $41.59$ & $8.73$ & 0.43 & $594.80$ & $\; \; 0.96$ \\
\hline \hline
\end{tabular}
\normalsize
\medskip
\caption{$\chi^{2}$ values of the best fit holographic model
($\Omega_{X}=0.707 \pm 0.009$, $\; c^{2}=0.407^{+0.033}_{-
0.028}$,$\; r_{f}=0.013^{+0.006}_{- 0.005}$, and $H_{0} = 71.8 \pm
2.9\,$km/s/Mpc), and the best fit $\Lambda$CDM model ($
\Omega_{M0} = 0.266 \pm 0.006$, and $H_{0}=71.8 \pm 1.9\,$
km/s/Mpc).} \label{table:chi2}
\end{table}
%%%%%%%%%%%%%%%%%%%%%%%%%%%%%%%%%%%%%%%%%%%%%%%%%%%%%%%%%%%%%%%%%%%%%%%
%%%%%%%%%%%%%%%%%%%%%%%%%%%%%%%%%%%%%%%%%%%%%%%%%%%%%%%%%%%%%%%%%%%%%%

Although the $\Lambda$CDM model fits the data somewhat better,
$\Delta \chi^{2} \simeq 2.5$, than the holographic model, that has
two more free parameters, it its uncertain which model should be
preferred in view that the former cannot address the cosmic
coincidence problem and the latter substantially alleviates it.
More abundant and accurate data, especially at redshifts between
the supernovae range and the CMB, will help decide the issue.
Nonetheless, we believe that the uncertainty will likely persists
till a breakthrough on the theoretical side allows us to calculate
with confidence the true value of the cosmological constant.

%%%%%%%%%%%%%%%%%%%%%%%%%%%%%%%%%%%%%%%%%%%%%%%%%%%%%%%%%%%%%%%%%%%%%%%%%%%%%%%%%%%%%%%%%%%%%%%%%%%%%
%%%%%%%%%%%%%%%%%%%%%%%%%%%%%%%%%%%%%%%%%%%%%%%%%%%%%%%%%%%%%%%%%%%%%%%%%%%%%%%%%%%%%%%%%%%%%%%%%%%%%
%%%%%%%%%%%%%%%%%%%%%%%%%%%%%%%%%%%%%%%%%%%%%%%%%%%%%%%%%%%%%%%%%%%%%%%%%%%%%%%%%%%%%%%%%%%%%%%%%%%%%

\section{Concluding remarks}\label{conclusions}
We performed a statistical study of the best fit parameters of the
holographic model -presented in section \ref{themodel}- using data
from SN Ia, CMB-shift, BAO, x-ray, the Hubble history, and the
growth function; 621 data in total. The maximum likelihood (or
minimum $\chi^{2}$) parameters are $\Omega_{X0}=0.707 \pm 0.009$,
$c^{2}=0.407^{+0.033}_{-0.028}$, $r_{f}= 0.013^{+0.006}_{-0.005}$,
and $H_{0}=71.8\pm2.9$ km/s/Mpc with $\chi^{2}/dof \approx 0.96$.
The $\Omega_{X0}$ and $H_{0}$ values fall within $1\sigma$ of the
corresponding values determined by Komatsu {\it et al.}
\cite{komatsu} ($0.734 \pm 0.029$ and $71.0 \pm 2.5$ km/s/Mpc,
respectively).  The evolution of the equation of state parameter,
$w$, at redshift below $1.2$ (see right panel of Fig. \ref{fig:w})
is compatible with the observational constraints derived in
\cite{serra-prd} and, as in other Ricci's holographic models
\cite{gao,lxu,suwa,lepe}, it crosses the phantom divide line at
recent times (see right panel of Fig. \ref{fig:w}). Curiously
enough, the $\Omega_{X0}$ and $c^{2}$ best fit values of this
model agree within $1\sigma$ with the corresponding values
obtained by Suwa {\it et al.} \cite{suwa} despite the use of a
very different interaction term between the dark components.

It is to be emphasized that holographic models do not contain the
$\Lambda$CDM model as a limiting case since the energy density of
the quantum vacuum, being constant, cannot be holographic. It is
also noteworthy that, in general, the $c^{2}$ term in the
holographic expression for the dark energy, Eq. (\ref{rhox}),
should not be considered constant except precisely when the
Ricci's length is chosen as the infrared cutoff \cite{nd-jcap}.

A lingering problem, both for this model and the $\Lambda$CDM
model, refers to the age of the old quasar APM 08279+5255 at
redshift $z = 3.91$. In both models the measured quasar age would
fall within $1\sigma$ only if the Hubble constant, $H_{0}$, would
come down substantially -something we do not expect though it
cannot be excluded. Accordingly, we must wait for further
observational data to see whether the present tension gets
exacerbated or disappears.

Before closing, it results interesting to contrast the model
explored in this paper with the model of Ref. \cite{id-jcap}. The
present one shows a better fit to the growth function at low
redshifts, as well as to the CMB-shift and BAO data. However, it
does not fit so well the age of the old quasar APM 08279+5255.

%%%%%%%%%%%%%%%%%%%%%%%%%%%%%%%%%%%%%%%%%%%%%%%%%%%%%%%%%%%%%%%%%%%%%%%%%%%%%%%%%%%%%%%%%%%%%%%%%%%%%%%%%%
%%%%%%%%%%%%%%%%%%%%%%%%%%%%%%%%%%%%%%%%%%%%%%%%%%%%%%%%%%%%%%%%%%%%%%%%%%%%%%%%%%%%%%%%%%%%%%%%%%%%%%%%%%
\acknowledgments{Thanks are due to Fernando Atrio-Barandela and
Luis Chimento for comments and advice on a earlier draft of this
work. ID was funded by the ``Universidad Aut\'{o}noma de
Barcelona" through a PIF fellowship. This research was partly
supported by the Spanish Ministry of Science and Innovation under
Grant FIS2009-13370-C02-01, and the ``Direcci\'{o} de Recerca de
la Generalitat" under Grant 2009SGR-00164.}

%%%%%%%%%%%%%%%%%%%%%%%%%%%%%%%%%%%%%%%%%%%%%%%%%%%%%%%%%%%%%%%%%%%%%%%%%%%%%%%%%%%%%%%%%%%%%%%%%%%%%%%%%%
%%%%%%%%%%%%%%%%%%%%%%%%%%%%%%%%%%%%%%%%%%%%%%%%%%%%%%%%%%%%%%%%%%%%%%%%%%%%%%%%%%%%%%%%%%%%%%%%%%%%%%%%%%


\begin{thebibliography}{99}
\bibitem{recentreviews}
J.A. Friemann, M.S. Turner and D. Huterer, Ann. Rev. Astron.
Astrophys. \textbf{46}, 385 (2008); R. Durrer and R. Maartens,
Gen. Relativ. Grav. \textbf{40}, 301; L. Amendola and Tsujikawa,
{\em Dark Energy. Theory and Observations} (CUP, Cambridge, 2010).
\bibitem{gerard-leonard}G. 't Hooft, ``Dimensional
reduction in quantum gravity", preprint gr-qc/9310026; L.
Susskind, J. Math. Phys. (N.Y.) \textbf{36}, 6377 (1995).
\bibitem{cohen}A. G. Cohen, D.B. Kaplan and A.E. Nelson, Phys.
Rev. Lett. \underline{82}, 4971 (1999); M. Li, Phys. Lett. B
\underline{603}, 1 (2004).
\bibitem{jng} Y.J. Ng, Phys. Rev. Lett. \underline{86}, 2946
(2001); M. Arzano, T.W. Kephart, and Y.J. Ng, Phys. Lett. B
\underline{649}, 243 (2007).
\bibitem{cqg-wd} W. Zimdahl and D. Pav\'{o}n, Class. Quantum Grav.
\underline{24}, 5461 (2007).
\bibitem{hubbleradius} S. D. H. Hsu, Phys. Lett. B
\underline{594}, 13 (2004); D. Pav\'{o}n and W. Zimdahl, Phys.
Lett. B \underline{628}, 206 (2005); B. Guberina, R. Horvat, and
H. Nikolic, JCAP01 (2007) 012; L. Xu, JCAP09 (2009) 016.
\bibitem{gao} C. Gao, F. Wu, X. Chen, and Y.G. Shen, Phys. Rev. D
\underline{79}, 043511 (2009).
\bibitem{lxu}
L. Xu, W. Li, and J. Lu, Mod. Phys. Lett. A \underline{24}, 1355
(2009).
\bibitem{suwa}
M. Suwa, T. Nihei, Phys. Rev. D \underline{81},
023519 (2010).
\bibitem{lepe}
S. Lepe and F. Pena, European Physical Journal (in the press),
arXiv:10052.2180 [hepth].
\bibitem {brustein} R. Brustein, ``Cosmological entropy bounds", in
{\em String Theories and Fundamental Interactions}, Lecture Notes
Physics \underline{737}, 619 (2008), hep-th/0702108.
\bibitem{luca} L. Amendola,  Phys. Rev. D \underline{62}, 043511 (2000);
D. Tocchini-Valentini and  L. Amendola, Phys. Rev. D, 65 063508
(2002).
\bibitem{srd} S. del Campo, R. Herrera, and D. Pav\'{o}n,
Phys. Rev. D \underline{78}, 021302(R) (2008); {\em ibid.} JCAP01
(2009) 020.
\bibitem{abdalla-abramo} E. Abdalla, L.R. Abramo, and J.C.C.
Souza, Phys. Rev. D \underline{82}, 023508 (2010).
\bibitem{lee}
J. Lee, Astrophys. J. Letters (in the press), arXiv:1008.4620
[astro-ph.CO].
\bibitem{wetterich} C. Wetterich, Nucl. Phys. B, \underline{302}, 668
(1988);{\it ibid} Astron. Astrophys. \underline{301}, 321 (1995).
\bibitem{jerome} Jer\^{o}me Martin, personal communication.
\bibitem{prd-lad} L.P. Chimento, A.S. Jakubi, and D. Pav\'{o}n,
Phys. Rev. D \underline{67}, 087302 (2003).
\bibitem{ladw} L.P. Chimento, A.S. Jakubi, D. Pav\'{o}n, and W.
Zimdahl, Phys. Rev. D., Phys. Rev. D \underline{67}, 083513
(2003); G. Olivares, F. Atrio-Barandela, and D. Pav\'{o}n, Phys.
Rev. D \underline{71}, 063523 (2005).
\bibitem{db-grg} D. Pav\'{o}n and B. Wang, Gen. Relativ. Grav.
\underline{41}, 1 (2009).
\bibitem{serra-prd}
P. Serra \textit{et al.}, Phys. Rev. D \underline{80}, 121302
(2009).
\bibitem {Daly} R.A. Daly, S.G. Djorgovski, K.A. Freeman, M. Mory,
C.P. O'Dea, P. Kharb, and S. Baum, Astrophys. J. \textbf{677}, 1
(2008).
\bibitem{id-jcap} I. Dur\'{a}n, D. Pav\'{o}n, and W. Zimdahl, JCAP07 (2010)
018.
\bibitem{ademir} A.C.S. Fria\c{c}a, J.S. Alcaniz, and J.A.S. Lima,
Mon. Not. R. Astron. Soc. \underline{362}, 1295 (2005).
\bibitem{dunlop_1999} J. Dunlop \textit{et al.}, ``Old Stellar Populations in
Distant Radio Galaxies", in {\em The Most Distant Radio Galaxies},
eds. H.J.A. Rottgering, P. Best and M.D. Lehnert (Kluwer,
Dordrecht, 1999), p. 71.
\bibitem{dunlop_1996} J. Dunlop \textit{et al.}, Nature (London)
\underline{381}, 581 (1996).
\bibitem{spinrad} H. Spinrad \textit{et al.}, Astrophys. J.
\underline{484}, 581 (1997).
\bibitem{hasinger} G. Hasinger \textit{et al.}, Astrophys. J.
\underline{573}, L77 (2002); S. Komossa and G. Hasinger, in Proc.
of the Workshop ``XEUS -Studying the Evolution of the Hot
Universe", eds. G. Hasinger  \textit{et al.}, astro-ph/0207321.
\bibitem {Amanullah} R. Amanullah \textit{et al.} (The Supernova Cosmology Project),
Astrophys. J. (in the press), arXiv:1004.1711.
\bibitem {Nesseris-Perilovorapoulos_0} S. Nesseris and L. Perivolaropoulos,
Phys. Rev. D \underline{72}, 123519 (2005).
\bibitem{wang-mukherjee} Y. Wang and P. Mukherjee, Astrophys. J.
\underline{650}, 1 (2006).\bibitem{bond} J.R. Bond, G. Efstathiou,
and M. Tegmark, Mon. Not. R. Astron. Soc. \underline{291}, L33
(1997).
\bibitem {komatsu} E. Komatsu \textit{et al.}, Astrophys.
J. Suppl. \underline{180}, 330 (2009).
\bibitem {Eisenstein} D.J. Eisenstein \textit{et al.} [DSS Collaboration]
Astrophys. J. \underline{633}, 560 (2005).
\bibitem {Percival} W.J. Percival \textit{et al.}, Mon. Not. R. Astron. Soc.
\underline{401}, 2148 (2010).
\bibitem{Nesseris-Perilovorapoulos-jcap} S. Nesseris and L. Perivolaropoulos,
JCAP01 (2007) 018.
\bibitem{White} S.M.D. White, J. F. Navarro, A. Evrard, and
C.S. Frenk, Nature \underline{366}, 429 (1993).
\bibitem {Allen} S.W. Allen \textit{et al.}, Mon. Not. R. Astron. Soc.
\underline{383}, 879  (2008).
\bibitem{Riess-2009}A.G. Riess, {\em et al.}, Astrophys. J.
\underline{699}, 539 (2009).
\bibitem{gazta} E. Gazta\~{n}aga, A. Cabr\'{e}, and L. Hui,
Mon. Not. R. Astron. Soc. \underline{399}, 1663 (2009).
\bibitem{simon} J. Simon, L. Verde, and R. Jim\'{e}nez, Phys. Rev. D
\underline{71}, 123001 (2005).
\bibitem{stern} D. Stern, R. Jim\'{e}nez, L. Verde, M.
Kamionkowski, and S.A. Stanford, JCAP02 (2010) 008.
\bibitem{Gong} Y. Gong, Phys. Rev. D \underline{78}, 123010 (2008).
\bibitem{nd-jcap} N. Radicella and D. Pav\'{o}n, JCAP10 (2010)
005.
\end{thebibliography}
\end{document}